\documentclass[iop,apj,tighten]{emulateapj}
\usepackage{graphicx,enumitem}

\shorttitle{A VLT/FLAMES study of NGC 1846 -- I. Kinematics}
\shortauthors{Mackey et al.}

\begin{document}

%% LaTeX will automatically break titles if they run longer than
%% one line. However, you may use \\ to force a line break if
%% you desire.

\title{A VLT/FLAMES study of the peculiar intermediate-age Large Magellanic Cloud\\star cluster NGC 1846 -- I. Kinematics\altaffilmark{$\dagger$}}

%% Use \author, \affil, and the \and command to format
%% author and affiliation information.
%% Note that \email has replaced the old \authoremail command
%% from AASTeX v4.0. You can use \email to mark an email address
%% anywhere in the paper, not just in the front matter.
%% As in the title, use \\ to force line breaks.

\author{A. D. Mackey$^1$, G. S. Da Costa$^1$, A. M. N. Ferguson$^2$, and D. Yong$^1$}
\affil{$^1$\ Research School of Astronomy \& Astrophysics, The Australian National University, Mount Stromlo 
Observatory,\\via Cotter Road, Weston, ACT 2611, Australia; \url{dougal@mso.anu.edu.au}\\
$^2$\ Institute for Astronomy, University of Edinburgh, Royal Observatory, Blackford Hill, Edinburgh, EH9 3HJ, UK.}

\altaffiltext{$\dagger$}{Based on observations obtained at the European Southern
Observatory Very Large Telescope, Paranal, Chile, under programme 082.D-0387.}

%% Mark off your abstract in the ``abstract'' environment. In the manuscript
%% style, abstract will output a Received/Accepted line after the
%% title and affiliation information. No date will appear since the author
%% does not have this information. The dates will be filled in by the
%% editorial office after submission.

\begin{abstract}
In this paper we present high resolution VLT/FLAMES observations of red giant stars in the massive 
intermediate-age Large Magellanic Cloud star cluster NGC 1846, which, on the basis of its extended 
main-sequence turn-off (EMSTO), possesses an internal age 
spread of $\approx 300$ Myr. We describe in detail our target selection and data reduction procedures,
and construct a sample of $21$ stars possessing radial velocities indicating their membership of
NGC 1846 at high confidence. We consider high-resolution spectra of the planetary nebula Mo-17, and 
conclude that this object is also a member of the cluster. Our measured radial velocities allow us to 
conduct a detailed investigation of the internal kinematics of NGC 1846, the first time this has been
done for an EMSTO system. The key result of this work is that the cluster exhibits a significant degree 
of systemic rotation, of a magnitude comparable to the
mean velocity dispersion. Using an extensive suite of Monte Carlo models we demonstrate that,
despite our relatively small sample size and the substantial fraction of unresolved binary stars in
the cluster, the rotation signal we detect is very likely to be genuine. Our observations are in qualitative
agreement with the predictions of simulations modeling the formation of multiple populations of
stars in globular clusters, where a dynamically cold, rapidly rotating second generation is a common
feature. NGC 1846 is less than one relaxation time old, so any dynamical 
signatures encoded during its formation ought to remain present.
\end{abstract}

%% Keywords should appear after the \end{abstract} command. The uncommented
%% example has been keyed in ApJ style. See the instructions to authors
%% for the journal to which you are submitting your paper to determine
%% what keyword punctuation is appropriate.

\keywords{globular clusters: individual: NGC 1846 --- Magellanic Clouds --- stars: kinematics and dynamics}

%% From the front matter, we move on to the body of the paper.
%% In the first two sections, notice the use of the natbib \citep
%% and \citet commands to identify citations.  The citations are
%% tied to the reference list via symbolic KEYs. The KEY corresponds
%% to the KEY in the \bibitem in the reference list below. We have
%% chosen the first three characters of the first author's name plus
%% the last two numeral of the year of publication as our KEY for
%% each reference.

\section{Introduction}
\subsection{Multiple populations in Galactic globular clusters}
One of the outstanding problems in modern astrophysics concerns the formation of globular clusters.
Long thought to constitute a simple, homogeneous class of object, each one consisting of stars of a 
uniform age and elemental composition, it is now recognised that these systems harbor multiple stellar
populations displaying a wide variety of unexpected characteristics \citep[see][for a review]{gratton:12}.

All Galactic globular clusters 
for which large samples of members have been studied spectroscopically at sufficiently high resolution and 
signal-to-noise are found to be comprised of stars exhibiting a characteristic chemical signature -- a strong 
anti-correlation between the abundances of the light elements O-Na, as well as C-N and Mg-Al in many cases, 
even while remaining homogeneous in iron content \citep[see e.g.,][]{carretta:09a,carretta:09b,carretta:09c}. 
Since this pattern is observed in stars on the main sequence as well as in red giants, it must be primordial rather 
than being the result of nucleosynthesis and mixing within the observed giant stars \citep[e.g.,][]{gratton:04}. 
The pattern is also seen in ancient globular clusters in nearby galaxies -- for example, in the Large Magellanic Cloud 
\citep{hill:00,johnson:06,mucciarelli:09} and the Fornax \citep{letarte:06} and Sagittarius 
\citep[e.g.,][]{carretta:10a,carretta:10b} dwarfs -- but not observed in old open clusters 
\citep[with the recent exception of NGC 6791,][]{geisler:12} or for the vast majority of the Milky Way field halo 
\citep[see e.g.,][and references therein]{martell:10,martell:11}, implying that it is a product specifically linked to 
globular cluster formation processes.

Beyond the anti-correlated light element abundance variations many globular clusters exhibit various additional levels 
of inhomogeneity, such as split main sequences or multiple sub-giant branches on their color-magnitude diagrams
(CMDs) \citep[e.g.,][]{bedin:04,piotto:07,villanova:07,milone:08}, internal dispersions in the abundance of iron
or other elements \citep[e.g.,][]{yong:08,dacosta:09,ferraro:09,cohen:10,mucciarelli:12}, and/or subpopulations 
enhanced in helium \citep[e.g.,][]{norris:04,piotto:05,dupree:11,pasquini:11}.

The overall picture is of a huge wealth of complexity that poses serious challenges for models of globular cluster
formation and evolution. The pervasive light element anti-correlations require material processed at high temperatures
via proton capture reactions. At $T \ga 2\times 10^7$\ K the CNO and NaNe cycles serve to alter the carbon, nitrogen, oxygen
and sodium abundances, while at somewhat higher temperatures the MgAl cycle also becomes active, leading to enhanced 
aluminium and reduced magnesium abundances. In the presently favored model, material processed in this way in a first
generation of stars pollutes or forms a central accumulation of gas in a young globular cluster, and a second generation
of stars bearing the characteristic light element signature is subsequently formed from this reservoir
\citep[see e.g.,][]{dercole:08,conroy:11}. Note that because the iron content in most clusters is observed to be
homogeneous, the gas should typically not have undergone supernova enrichment. Leading suggestions for
the sites of the high temperature processing are intermediate-mass asymptotic giant branch (AGB) stars \citep[e.g.,][]{ventura:01} 
and fast-rotating massive stars \citep[FRMS; e.g.,][]{decressin:07}. In both cases helium enhancement may also be introduced to the
gas reservoir as a result of main sequence hydrogen burning in the polluting stars. 

A generic prediction of this model
is that the second generation of stars in a globular cluster ought to be more centrally concentrated than the first generation, and
this seems to have been observed in some cases \citep{lardo:11}. However, the model also suffers from a number of difficulties, 
chief among which is that the 
second generation stars in globular clusters require significantly more gas for formation than can have been available based 
on the presently-observed numbers of first generation members. Ideas invoked to circumvent this issue include the 
accretion of large amounts of pristine interstellar material \citep[e.g.,][]{conroy:11}, a first generation $\sim 10-100$ 
times more massive than present-day globular clusters \citep[e.g.,][]{dercole:08,bekki:11}, a first generation with a top-heavy 
initial mass function \citep[e.g.,][]{dantona:04,bekki:06}, or a variation in which globular clusters are formed as part of initially 
larger systems such as low-mass dwarfs \citep[e.g.,][]{bekki:07}. Beyond this difficulty, it is 
also unclear how clusters with internal dispersions in iron-peak elements, or that include populations with very high 
helium abundances ($Y\approx 0.4$) fit into the model.

The unavoidable conclusion from the scenario outlined above is that the formation of individual globular clusters must
have spanned a period of tens to hundreds of Myr, depending on the nature of the stars responsible for the high temperature
processing. With presently available facilities, however, we are unable to directly resolve age differences of this magnitude 
given that Galactic globular clusters are typically $\sim 12$ Gyr old. In addition, since the time-scale for dynamical relaxation
in the majority of Galactic globular clusters is much shorter than the cluster age \citep[see][]{harris:96}
any detailed information imprinted on the internal kinematics of these systems as a result of the formation process 
will have long since been seriously diluted or possibly erased altogether.

\subsection{The role of peculiar Magellanic Cloud clusters}
Star clusters in the Large and Small Magellanic Clouds (LMC and SMC) offer an important new piece of this puzzle. 
These two galaxies possess extensive systems of clusters spanning the full age range $\sim 10^6 - 10^{10}$ years; 
many of the members of these systems are comparable in mass to present-day Galactic globular clusters lying at or 
below the peak of the luminosity function \citep[see e.g.,][]{mackey:03a,mackey:03b}. Using images taken with the 
Advanced Camera for Surveys (ACS) onboard the Hubble Space Telescope (HST), we recently demonstrated that several 
rich intermediate-age clusters ($\tau \sim 1.5-2$ Gyr) in the LMC display markedly unusual CMDs \citep{mackey:07,mackey:08a}. 
More specifically, we found that while the CMDs for NGC 1783, 1806 and 1846 have very narrow main sequences and 
red giant branches, the main-sequence turn-offs for these clusters are much broader than can be explained by the 
photometric uncertainties %, and, at least for NGC 1806 and 1846, may even be bifurcated 
(see Figure \ref{f:cmds}). 
After considering various possibilities for the origin of these extended main-sequence turn-offs (EMSTOs), such
as confusion due to unresolved binary stars or field contamination, we concluded that the simplest viable interpretation 
of our observations is that each of the three clusters is comprised of two or more stellar populations with very similar 
iron abundance but spanning an age interval of $\approx 300$ Myr. 

Subsequent work has reinforced this interpretation and revealed that an EMSTO is apparently not an
unusual feature for intermediate-age Magellanic Cloud clusters. Photometric analysis of HST imaging by \citet{milone:09}
and \citet{goudfrooij:09,goudfrooij:11a} demonstrated that, of the $16$ intermediate-age LMC clusters with suitable
data, $11$ possess EMSTOs \citep[see also][]{mackey:09}. In addition, \citet{glatt:08} discovered that the intermediate-age 
SMC cluster NGC 419 possesses an EMSTO. Each of these studies concluded that an internal age spread was the most 
likely explanation for these features, with the full sample of EMSTO clusters encompassing the range $\sim 150-500$ Myr 
for this spread. 

Additional support has come from several directions. \citet{girardi:09} noted the presence of a dual red clump in 
NGC 419 and in a number of the rich LMC EMSTO clusters, while \citet{rubele:10,rubele:11} used the complete 
observed CMDs for NGC 419 and NGC 1751 to reconstruct their star-formation histories -- finding that these may have 
spanned an incredible $\sim 700$ Myr (out of a mean age of $\sim 1.5$ Gyr) in NGC 419, and $\sim 460$ Myr in NGC 1751. 
\citet{rubele:10,rubele:11} noted that in their best-fitting models the dual red clumps in both NGC 419 and NGC 1751 arise 
as a direct result of the age dispersions in these systems. \citet{bastian:09}
posited that rather than reflecting an internal age spread, a cluster EMSTO might instead arise if a wide range
in stellar rotation is present at the MSTO. However, models by \citet{girardi:11} demonstrated that the effects of stellar rotation 
fail to reproduce the EMSTO morphology; in addition, as emphasised by \citet{rubele:10,rubele:11}, this scenario does not seem 
able to account for the dual red clumps observed in NGC 419, NGC 1751, and some other EMSTO clusters. \citet{girardi:11} also 
tested the effects of star-to-star variations in the degree of convective overshooting on a cluster CMD, but found that this too 
failed to accurately reproduce the characteristic EMSTO shape. 

The global properties of EMSTO clusters provide some clues as to how these systems may arise. \citet{conroy:11} made
the simple observation that it is only the intermediate-age clusters with present-day masses greater than 
$\approx 10^4\,{\rm M}_\odot$ that exhibit EMSTOs. \citet{keller:11} further observed that the known EMSTO systems
are, without exception, the most diffuse, spatially extended clusters for their age. In the framework of \citet{elson:87}
and \citet{mackey:03a,mackey:03b}, Magellanic Cloud clusters exhibit an increasing spread in size (defined by either
the core or half-light radius) with age; \citet{keller:11} showed that at intermediate ages the EMSTO clusters all fall
towards the upper envelope of this distribution, while clusters without an EMSTO appear systematically more compact.
\citet{mackey:08b} used detailed $N$-body models to explore the origin of the increasing spread in Magellanic Cloud
cluster sizes with age. They found that if very young massive clusters are formed as compact dense systems, as is 
seen to be the case in both the LMC and SMC, the only viable way for them to evolve along the upper envelope of the 
observed age-size distribution on a $\la 1$ Gyr time-scale is due to mass-loss from stellar evolution {\it if they 
were initially highly mass segregated} -- that 
is, with the highest mass stars preferentially located towards the cluster centers\footnote{\citet{mackey:08b}
found that a retained population of stellar mass black holes can also cause a cluster to move towards the upper
envelope of the age-size distribution, but on longer time-scales than $1-2$ Gyr. Note also that \citet{elson:89} found
that a flat, or top-heavy IMF can lead to cluster expansion along the upper envelope of the age-size distribution; however 
such clusters rapidly become unbound (after just a few$\,\times 10^7$ yr). In addition, young Magellanic Cloud clusters 
are seen to have quite normal IMFs \citep[e.g.,][]{kroupa:01,degrijs:02}.}. 
This in turn implies that, in addition to being the most massive intermediate-age clusters in the Magellanic Clouds,
EMSTO systems were also probably the most strongly mass segregated clusters at early times. There is clearly a link
between cluster mass and structure, and the presence of an EMSTO.

There are two additional observations of relevance. If a cut is made on the CMD across a cluster's EMSTO in a direction 
perpendicular to the locus of the upper main sequence, then groups of ``younger'' and ``older'' MSTO stars may be defined. 
%\footnote{For completeness we note that there is some disagreement about the precise structure of the EMSTO
%region in the richest clusters where this feature is observed. For example, \citet{mackey:07} and \citet{mackey:08a} found 
%the EMSTOs in NGC 1846 and 1806 to be split or bifurcated, but that in NGC 1783 to be more evenly spread. In contrast,
%\citet{milone:09} found all three to be bifurcated, whereas \citet{goudfrooij:09} and \citet{goudfrooij:11a} think they are
%all evenly spread. Irrespective of this detail, it is trivial to split the EMSTO in a cluster into ``younger'' and ``older'' groups.}. 
These groups have distinct properties. First, as noted by \citet{milone:09}, 
the younger population consists of at least as many stars as the older population -- in fact the ratio is typically more like 
$2$:$1$ in the richest clusters. Second, as described in detail by \citet{goudfrooij:11b}, the younger population is often 
more centrally concentrated than the older population. The strongest difference between the two concentrations is seen 
in the clusters possessing the largest estimated {\it initial} escape velocities (i.e., $v_{\rm esc} \ga 15$\ km$\,$s$^{-1}$
at an age of $\sim 10$ Myr); systems in which $v_{\rm esc} \la 10$\ km$\,$s$^{-1}$ do not show this difference in
central concentrations, or are not seen to possess an EMSTO at all (note that the limiting velocity here is comparable to
the velocities of winds from FRMS or AGB stars). This picture is fully consistent with the idea
discussed above that EMSTO clusters were both the most massive and the most strongly mass segregated clusters 
at the time when they were formed.

As speculated by a number of authors \citep{conroy:11,goudfrooij:11b,keller:11}, the properties of 
EMSTO systems suggest a formation process remarkably similar to that inferred for the
multi-population Galactic globular clusters -- specifically, that prolonged star formation has occurred at the centres 
of these objects because their masses and initial structures allowed the retention or accumulation of a suitable reservoir 
of gas at the bottom of the cluster potential well. The observed internal age spreads of several hundred Myr in EMSTO 
clusters, and the apparently minimal dispersions in iron abundance inferred from their narrow RGB sequences
are both consistent with the scenario invoked for Galactic globular clusters in which much of this reservoir comes 
from the slow winds of a first generation of AGB stars. Furthermore, the number ratio of younger to older groups of 
stars in EMSTO clusters is comparable to that seen for the two generations in Galactic globular clusters. 
Note that it has not been clearly assessed how early ideas for the formation of EMSTO systems, such as the merging
of two bound clusters \citep[e.g.,][]{mackey:07} or the merger of a cluster and a giant molecular cloud \citep{bekki:09},
fit with the observed properties of EMSTO systems. However, if the star formation has indeed progressed unbroken
over several hundred Myr as suggested by \citet{rubele:10,rubele:11} or \citet{goudfrooij:11a}, these scenarios are likely disfavored.

If, as hypothesized, EMSTO clusters and Galactic globular clusters share a common formation process, a key 
prediction is that EMSTO clusters ought to harbor similar star-to-star variations in the abundances of light elements
as seen in the Galactic systems. A complicating factor is the comparatively high overall metallicity of the EMSTO clusters 
($[$M$/$H$]\approx -0.4$), along with the possibility that their central gas resevoirs may have been augmented by an 
unknown amount of accreted pristine material \citep[e.g.,][]{conroy:11} -- so the scale of these variations is difficult to 
predict. Very few light element abundance measurements exist for EMSTO systems. The most extensive study is that by 
\citet{mucciarelli:08} who targeted between $5$ and $11$ stars in each of four intermediate-age LMC clusters, of which 
only one (NGC 1783) unambiguously features an EMSTO. For all four systems, Mucciarelli et al. assert that the star-to-star
scatter in each of the $\sim 20$ elemental abundances they measure is negligible; in particular, there are no O-Na or
Mg-Al anti-correlations evident. However, the interpretation of these measurements is somewhat controversial --
both \citet{conroy:11b} and \citet{goudfrooij:11b} note that the star-to-star scatter in the listed abundances of sodium 
is larger than the observational uncertainties at a statistically significant level (up to $\sim 4\sigma$) for some clusters. 
Even so, the sole EMSTO cluster in the study, NGC 1783, does not seem to exhibit elemental abundance patterns 
which are strikingly distinct from those of the other three systems.

\subsection{This work}
Additional high resolution spectral data, for a larger sample of stars in a larger ensemble of EMSTO clusters, are clearly 
required to assess the viability of the link between these systems and the multi-population Galactic globular clusters. 
Such data possess an extra benefit beyond exploring elemental abundance patterns. Because EMSTO clusters are diffuse, 
low-density stellar systems, their two-body relaxation times are typically comparable to, or longer than their ages 
\citep[see e.g.,][]{goudfrooij:11b}. Thus, unlike for Galactic globular clusters, the internal dynamics of EMSTO systems
should still reflect the conditions present early on in their lives -- so that any signature imparted onto the 
cluster kinematics by the formation process should be both detectable and straightforward to interpret.

This paper is the first in a series devoted to a detailed study of medium and high resolution spectra for $21$ giant
stars and one planetary nebula in the most massive known EMSTO cluster, and one of the best studied photometrically 
-- NGC 1846 in the LMC. Stellar populations in this cluster span the age range $\approx 1.6-1.9$ Gyr, and its 
iron abundance is $[$Fe$/$H$] \approx -0.4$ \citep[e.g.,][]{mackey:07,mackey:08a}. The aim of our study of NGC 1846
is two-fold: first, to characterise the internal dynamics of the cluster and search for any signatures of its 
formation that might be present; and second, to place constraints on any star-to-star elemental abundance variations, 
especially for light elements. Here, we present a detailed description 
of our target selection and data analysis (Sections \ref{s:data} and \ref{s:member}), and focus on the cluster kinematics
(Sections \ref{s:kinematics} and \ref{s:discuss}). Subsequent work will cover the elemental abundance analysis.

\section{Observations \& Data Reduction}
\label{s:data}
\subsection{Data Acquisition}
We obtained spectra of stars in the vicinity of NGC 1846 using the FLAMES instrument at the ESO
Very Large Telescope (VLT) on Cerro Paranal, Chile, under programme 082.D-0387 (PI: Mackey). 
FLAMES \citep{pasquini:02} is a fiber-fed  multi-object spectrograph mounted at the Naysmith A 
platform of the 8.2m Unit Telescope 2 (Kueyen). We employed the MEDUSA-GIRAFFE mode, allowing 
up to $132$ stars to be targeted across the $25\arcmin$ diameter field of view in a single pointing.

Our target selection is discussed in detail in Section \ref{ss:targets}. We used just a single FLAMES
fiber configuration but observed at three high-resolution GIRAFFE settings (HR11, HR13, and HR14B) and
one low-resolution setting (LR02). The nominal wavelength coverage and spectral resolution for each of
these set-ups is listed in Table \ref{t:flames}. This paper and the next in the series (Paper II) are dedicated 
to analysis of the observations obtained at the three HR settings; results from the blue LR set-up will be reported 
in a separate future work.

We obtained our data in visitor mode on the three nights 2008 November 29 -- December 01. 
Conditions were clear and stable, with seeing typically in the range $0.5\arcsec - 1.0\arcsec$.
On each night we observed a given HR setting at the beginning and
end of the night, and reserved an hour either side of NGC 1846 crossing the meridian for
observing the blue LR set-up with minimal differential atmospheric refraction. For all four gratings we 
obtained $6\times55$\ min exposures. Due to the faintness of our targets we switched the simultaneous 
calibration lamps off, instead bracketing each long science exposure with short calibration lamp exposures 
to ensure we could achieve an accurate wavelength solution. 

\subsection{Target Selection and Photometry}
\label{ss:targets}
Targets for our FLAMES observations were drawn from the archival HST/ACS imaging of NGC 1846 described
in the introduction \citep{mackey:07,goudfrooij:09} for the crowded central regions, and the Magellanic Clouds 
Photometric Survey (MCPS) catalogue of the LMC \citep{zaritsky:04} for the surrounding field. 

\begin{figure*}
\begin{center}
\includegraphics[width=180mm]{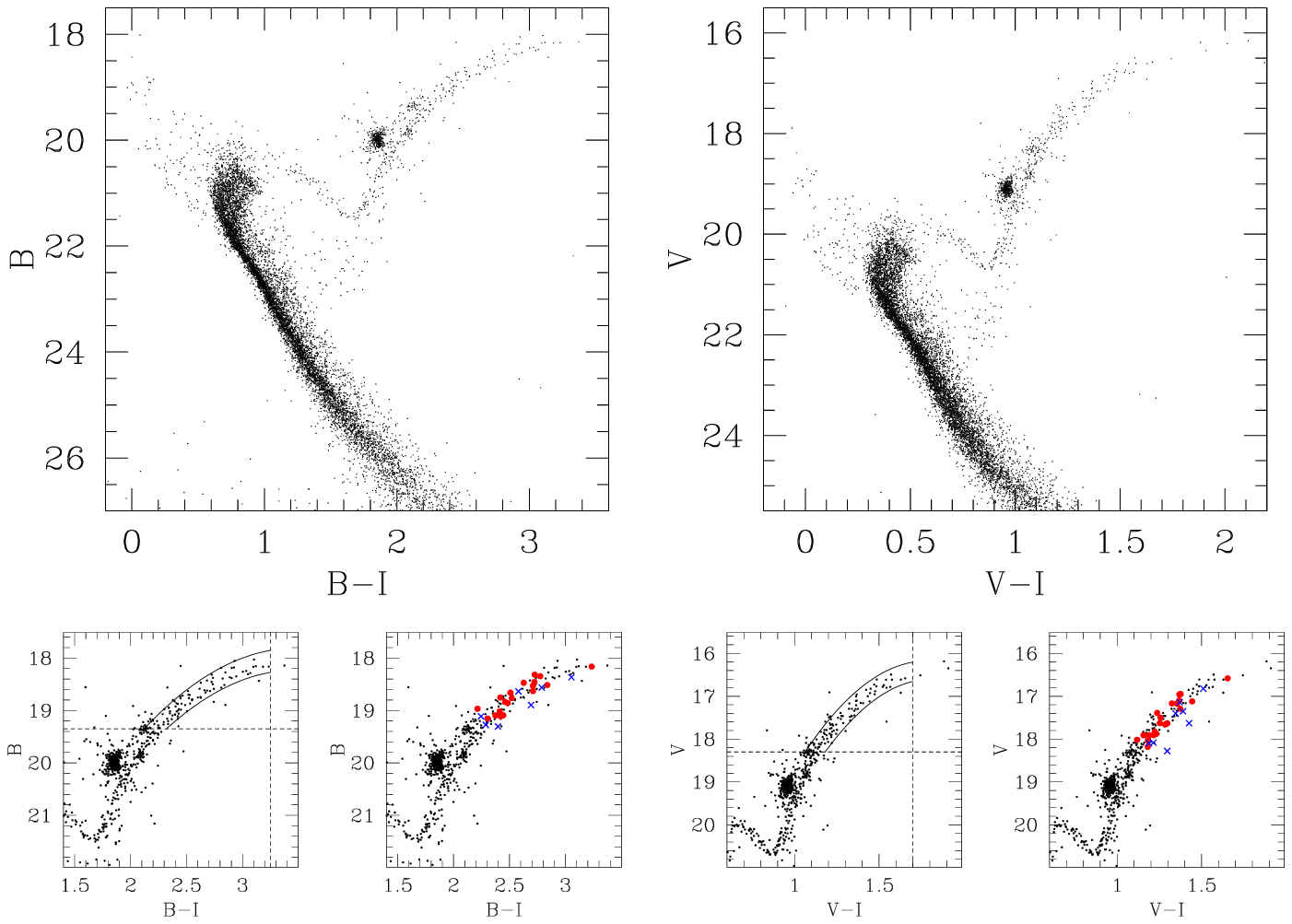}
\end{center}
\caption{HST/ACS color-magnitude diagrams for NGC 1846, in the $(B-I,\,B)$ and $(V-I,\,V)$ planes. In the upper two panels, the peculiar extended main-sequence turn-off morphology is clearly evident. The lower four panels show the regions on the red giant branch for our FLAMES target selection, bounded by dashed and solid lines, along with the objects we observed. Those targets marked with red dots are radial velocity members of the cluster (see Section \ref{ss:memstar}); those marked with blue crosses are not.\label{f:cmds}}
\end{figure*}
 
For work such as this, one would ideally directly target stars spread across the cluster EMSTO so that elemental 
abundances and dynamical properties can be correlated directly against position on the CMD. Unfortunately, 
however, at $V\approx 20.5$, obtaining sufficient signal-to-noise for a high precision abundance analysis of 
NGC 1846 MSTO stars is beyond the capabilities of presently-available high-resolution spectrographs, even on 
the largest telescopes. Instead, we are forced to target the brightest giant stars in the cluster. There is, however,
no major disadvantage in doing so -- \citet{milone:09} demonstrated that the relative numbers of ``younger''
and ``older'' stars across the EMSTO are not too dissimilar (a ratio of $\approx 2$:$1$). This means that,
providing a sufficiently large ensemble of giant stars is observed, the full spread in population parameters
(age and, if they exist, elemental abundance variations) ought to be well sampled.

\begin{deluxetable}{ccccccc}
\tabletypesize{\scriptsize}
\tablecaption{Nominal wavelength coverage and spectral resolution for our four GIRAFFE set-ups.\label{t:flames}}
\tablehead{
\colhead{Configuration} & \hspace{1mm} & \colhead{$\lambda_{\rm start}$} & \colhead{$\lambda_{\rm end}$} & \colhead{$\Delta\lambda$} & \hspace{1mm} & \colhead{$R$} \\
\colhead{Name} & \colhead{} & \colhead{(\AA)} & \colhead{(\AA)} & \colhead{(\AA)} & \colhead{} & \colhead{}
}
\startdata
HR11   & & $5597$ & $5840$ & $243$ & & $24\,200$ \\
HR13   & & $6120$ & $6405$ & $285$ & & $22\,500$ \\
HR14B & & $6383$ & $6626$ & $243$ & & $28\,800$ \\
LR02   & & $3964$ & $4567$ & $603$ & & $6\,000$ 
\enddata
\end{deluxetable}
 
At the time we were preparing our VLT observations, HST imaging of NGC 1846 was available from only 
two separate programs: 9891 (PI: Gilmore) and 10595 (PI: Goudfrooij), both of which utilised the Advanced 
Camera for Surveys (ACS) Wide Field Channel (WFC). The former is a ``snapshot'' program,
where the imaging consists of just two exposures -- $300$s in the F555W filter and $200$s in 
the F814W filter. The second program comprises a more extensive set of observations -- three exposures
in each of the F435W, F555W, and F814W filters, where two of the exposures were long ($340$s each)
and one short ($90$s, $40$s, and $8$s in the three filters, respectively).
The two programs were observed with differing orientations, meaning that their footprints only partially overlap 
on the sky. In the interests of covering as much of the area around NGC 1846 as possible with high quality ACS 
imaging and photometry, and hence maximising the number of objects in our input catalogue for FLAMES target
selection, we used the {\sc multidrizzle} software \citep{koekemoer:03} to combine the complete set of 
F555W observations into a ``master''  reference image. This image increases the sky coverage near NGC 1846
by $\sim 30\%$ over what would have been available using only one of the HST programs.

We next used the {\sc dolphot} software \citep{dolphin:00}, and in particular its ACS module, to photometer 
all the available images. Details of this procedure may be found in \citet{mackey:07}. 
Briefly, {\sc dolphot} performs point-spread function (PSF) fitting photometry using model PSFs especially
tailored for the ACS camera. It works on images for which basic reduction steps have been applied
(bias and dark current subtraction, and flat-field division) but which have not been distortion corrected
(drizzled). The software can photometer multiple images in multiple filters simultaneously, matching
detections across images and deriving coordinates relative to an input reference frame -- in our case
the master drizzled F555W image. Output photometry is on the calibrated VEGAMAG scale of 
\citet{sirianni:05} and has been corrected for charge-transfer efficiency degredation. Where possible,
transformations are also made into the standard Johnson-Cousins system. To obtain a clean
list of stellar detections with high quality photometry we filtered our {\sc dolphot} measurements using 
the classification, sharpness and crowding parameters \citep[see][]{mackey:07,mackey:08a}. 

\begin{figure*}
\begin{center}
\includegraphics[width=86mm]{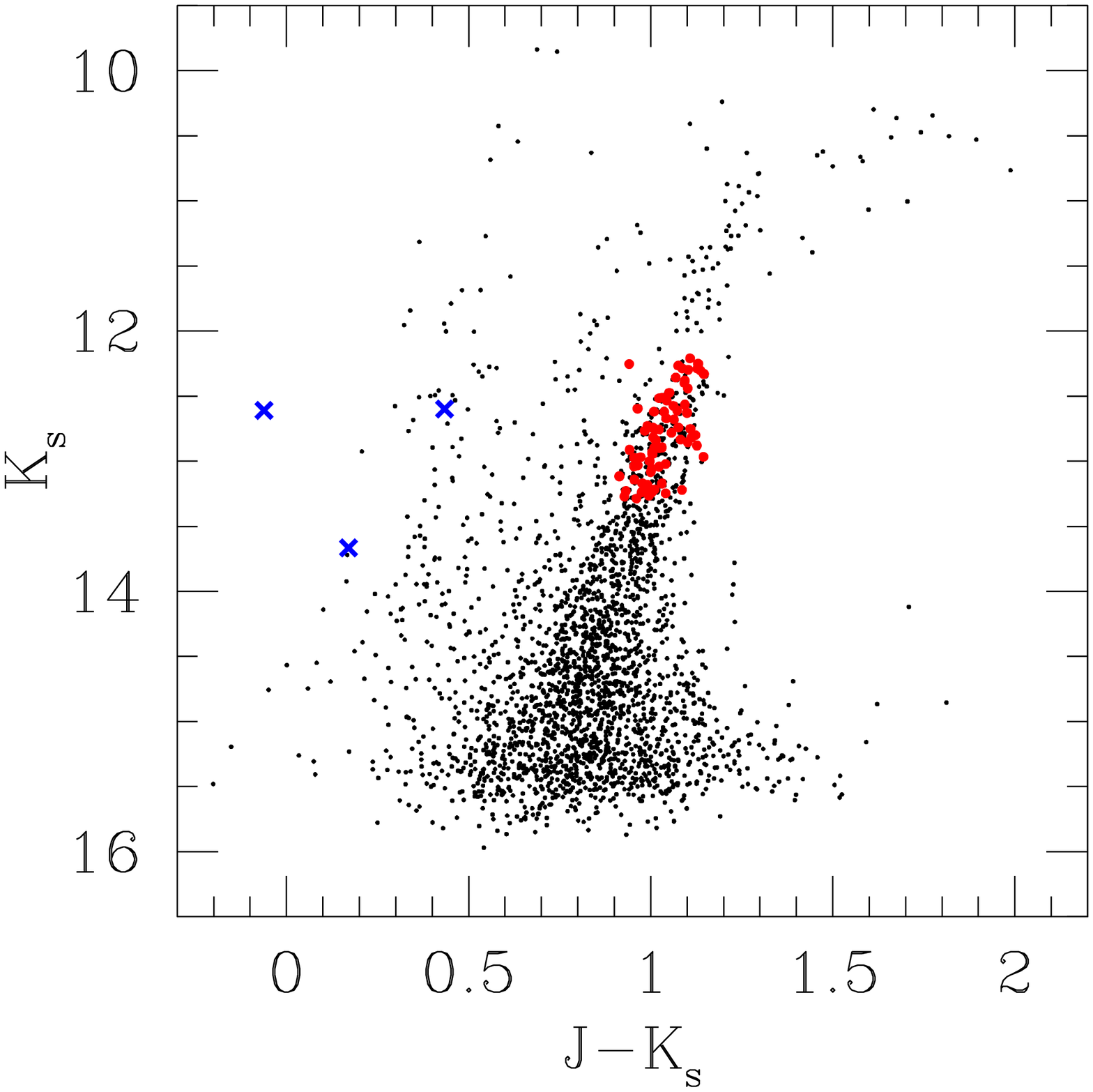} 
\hspace{-2mm}
\includegraphics[width=86mm]{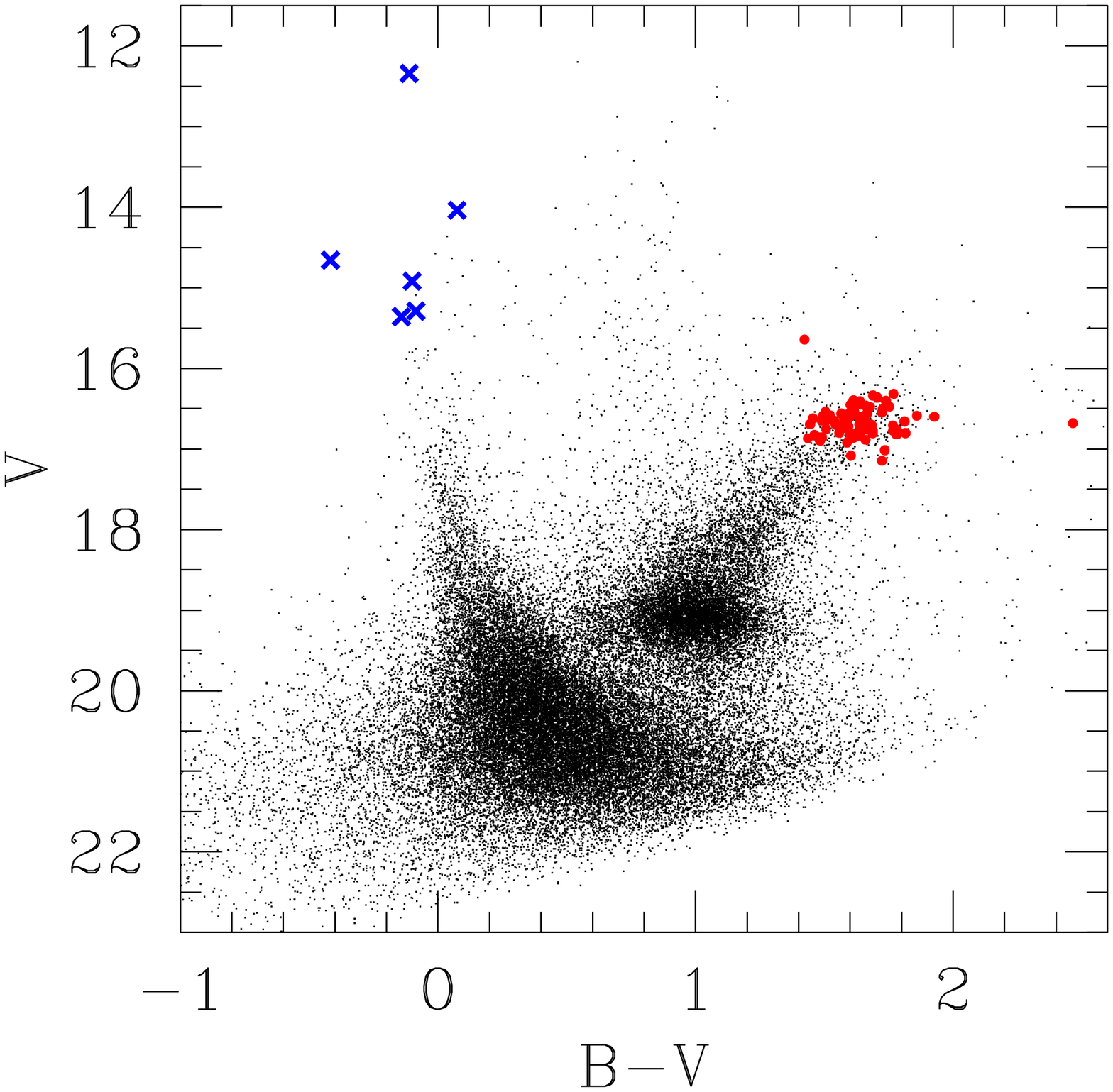}
\end{center}
\caption{Color-magnitude diagrams from the Magellanic Clouds Photometric Survey catalog \citep[see][]{zaritsky:04} for stars lying within the $25\arcmin$ diameter FLAMES field of view. Note that objects lying within the central $\approx 2\arcmin$, where NGC 1846 is located, have been excised. The left panel shows the $(J-K_s,\,K_s)$ plane, where the photometry originally comes from the 2MASS catalog, while the right panel shows the $(B-V,\,V)$ plane. Stars belonging to the LMC field for which we obtained FLAMES spectra are marked with red dots in both panels. Objects marked with blue crosses are our telluric correction stars (note that not all of these have infrared photometry).\label{f:mcps}}
\end{figure*}
 
CMDs, in both the $(B-I,\,B)$ and $(V-I,\,V)$ planes are shown in Figure \ref{f:cmds}. 
NGC 1846 is set against a relatively dense background of LMC field stars, so care was necessary
in selecting our input catalogue for FLAMES. We used CMDs for stars within $30\arcsec$ of the cluster
center to filter out most of the field contamination, allowing us to define regions encompassing the
upper red-giant branch (RGB) and asymptotic giant branch (AGB) on both colour-magnitude planes.
The colour and luminosity of a star on the RGB is sensitive to elemental abundances, especially that of 
iron, as well as to age. Given that the primary aim of our VLT program was to test for possible star-to-star 
abundance variations, and that we also strongly suspect the presence of an internal age spread in
NGC 1846, we were particularly mindful not to define our target regions too restrictively. The red side of
the NGC 1846 RGB is sharply defined on both CMD planes; however to the blue there is potential for
significant overlap between the RGB and AGB sequences. Hence we set the blue side of our target region
rather conservatively, with the result there are likely a few AGB stars in our FLAMES sample.
As long as these are early-AGB objects (i.e., which have not yet undergone third dredge-up) the composition
of the atmospheres of these stars should not have been altered substantially through the addition of
newly-processed material. This issue is not relevant to the cluster kinematics, but is discussed in more 
detail in Paper II.
%Nonetheless, we remained alert to this issue throughout our chemical abundance analysis.

We set an absolute red limit to our target region of $V-I = 1.7$ in order to exclude very cool giants
for which it is more difficult to do a reliable abundance analysis. Note that this also excluded the most luminous 
(and evolved) AGB stars in the cluster. At the faint end we set a limit of $V = 18.3$, as spectra
for stars fainter than this would not have sufficiently high signal-to-noise. On the $(B-I,\,B)$ plane,
the corresponding limits were $B-I = 3.25$ and $B = 19.35$.

Having defined our target regions we selected as our cluster input catalogue all stars across the full 
ACS field of view which lay within these regions on both colour-magnitude planes. As noted above, 
nearly a third of the field is not covered by F435W imaging -- we filtered objects in this area 
using only the $(V-I,\,V)$ plane. The very outer parts of the ACS field were only sparsely populated 
with suitable stars, so we supplemented the input catalogue in this region with stars lying up to $\sim 0.1$
mag to the red of our target region on the CMD. We also added to the catalogue a nearby planetary 
nebula likely belonging to NGC 1846, Mo-17 (see Section \ref{ss:pn}). We searched for, and removed 
from the catalogue any stars with neighbours within a radius of $2\arcsec$ that were sufficiently
bright as to be likely to interfere with the spectrum of the target.

We used ESO's Fibre Positioner Observation Support Software (FPOSS) to determine the optimal MEDUSA
configuration for our FLAMES observations. A few trial runs using our ACS input catalogue demonstrated
that we would be able to observe, at most, $\sim 30$ stars within $\approx 2\arcmin$ of the
center of NGC 1846 -- the limiting factor being the minimum fiber-to-fiber separation of $11\arcsec$.
Even allowing for $\sim 15$ sky fibers and $\sim 5$ fibers allocated to hot blue stars to allow correction 
of telluric absorption (see below), this left roughly $80$ fibers unused. Rather than waste these we
decided to allocate them to bright field RGB stars spread evenly over the non-cluster regions of the
$25\arcmin$ diameter FLAMES field of view.

We selected these objects from the Magellanic Clouds Photometric Survey (MCPS) catalogue of the LMC 
\citep[see][]{zaritsky:04}. This catalogue provides precise astrometry and $UBVI$ photometry of stars 
in the central $64$ deg$^2$ of the LMC, and, for some subset of successfully cross-matched stars,
$JHK_s$ photometry from the Two Micron All Sky Survey (2MASS) catalogue \citep{skrutskie:06}. There were a
sufficiently large number of MCPS stars with 2MASS matches across our field of view that we were able
to draw targets exclusively from this subset. The 2MASS measurements were particularly useful for 
distinguishing between field RGB and AGB stars, as shown in Figure \ref{f:mcps} (left panel). We selected 
RGB stars brighter than $V\approx 17$ for our MCPS input catalogue (Figure \ref{f:mcps}, right panel).

Available literature measurements suggested that the NGC 1846 radial velocity of $V_{\rm sys} \approx 240$ 
km$\,$s$^{-1}$ \citep{olszewski:91,grocholski:06} would bring the forbidden $[$O{\sc i}$]$ line at 
$6300.3\,$\AA\ into close proximity with telluric absorption features due to atmospheric O$_2$. We planned 
to derive our oxygen abundances using primarily this line (see Paper II); in order to be able to account for the telluric 
absorption we required spectra of hot, preferably fast-rotating stars at high S/N. To this end, we selected a 
number of target stars from the MCPS catalogue that appeared to be bright members of the young blue LMC 
field main sequence (see Figure \ref{f:mcps}).

To aid in the subtraction of atmospheric emission lines from our science spectra, we also decided to
allocate $\approx 15$ fibers to blank sky. We placed these such that for a given sky fiber there were no objects 
in the MCPS catalogue lying within $5\arcsec$. Ideally these sky positions would be local to our 
highest priority targets (i.e., NGC 1846 members); however the very tightly packed nature of the fiber configuration 
in the middle of the FLAMES field meant that it was impossible to allocate any other fibers within $\sim 5\arcmin$
of the cluster center. This ultimately led to some problems achieving a high quality sky subtraction for
many of our targets; however it was not difficult to accommodate these issues in our analysis (Section \ref{ss:reduction}).

Finally, we merged our ACS and MCPS catalogues for input to FPOSS. The critical step was to
transform the ACS coordinates onto the MCPS (FK5) astrometric frame. Although the relative astrometry
of targets measured from the ACS imaging is extremely precise (of order milli-arcseconds), the HST FITS 
header information from which the absolute astrometry is derived can be in error by up to several arcseconds
\citep[e.g.][]{anderson:08} -- more than enough for the MEDUSA fibres to miss these targets entirely.
We cross-matched $\sim 100$ stars across the ACS master reference image with stars in the full MCPS LMC
catalogue, excluding the crowded central cluster region, and used these to derive a suitable coordinate 
transformation onto the MCPS frame. 

Having successfully merged our two lists of potential targets, we used FPOSS to determine the optimal 
fiber configuration. In doing so, we assigned priorities to the various classes of target -- for example, ACS stars
received higher priority than MCPS stars and sky positions. The ACS stars themselves were graded in priority
by luminosity (the brighter the better), distance from the center of the cluster (the smaller the better), and, 
weakly, by colour on the RGB (the redder the better).

Our final configuration targeted $30$ ACS stars ($11$ within $50\arcsec$ of the cluster center), the planetary
nebula Mo-17, $79$ MCPS field RGB stars, $6$ MCPS stars for correcting telluric absorption, and $16$ blank sky 
positions. We observed our blue LR setting using FLAMES fiber positioner plate 1 and the three HR settings using 
plate 2. On both plates, spectra from two fibers (both targeting MCPS stars) fell off the edge of the CCDs and were 
lost. Additionally, two of the plate 2 fibers were broken leading to the loss of another MCPS star and one ACS 
target. Hence, the HR observations presented here resulted in spectra for $76$ MCPS stars and $29$
ACS stars, as well as the other targets and sky positions listed above.

\subsection{Data Reduction}
\label{ss:reduction}
We used the ESO public GIRAFFE pipeline recipes v2.8.1, operating under the graphical front-end software
{\sc gasgano}, to perform a basic reduction of all our science frames -- that is, bias subtraction, 
fiber localization, optimal extraction of spectra, division by a normalized flat-field image (i.e., including correction
for fiber-to-fiber transmission differences), an initial wavelength calibration, and rebinning to a uniform
linearized dispersion scale ($0.05\,$\AA\ per pixel for the spectra considered here). For each of the four
instrumental set-ups, the science frames included the six on-target exposures as well as the short bracketing 
exposures with the calibration lamps on. For science frames belonging to a given set-up, the initial wavelength
calibration was derived from an arc-lamp exposure taken at the beginning of the night on which the science
frames were taken. After the basic data reduction had been completed, we used the measured positions
of emission lines in the bracketing calibration frames to check for small residual wavelength drifts
in each of the on-target exposures individually. 

Next, we intended to perform a sky subtraction on each individual spectrum of an object, along with a 
correction for telluric absorption, before combining these spectra into a final product. We note that
sky subtraction is only marginally important for the HR11 set-up, 
which covers a handful of weak emission lines; however the HR13 and HR14B set-ups both cover numerous 
bright emission lines. Similarly, telluric absorption is negligible ($\la 3\%$) for the 
HR11 and HR14B set-ups, but is noticeable for the HR13 set-up over the range $6275-6330\,$\AA\ 
where there are many lines of $\sim 5-15\%$ absorption due to atmospheric O$_2$. 

Despite testing a variety of techniques we were unable to obtain a high-quality sky subtraction across the
full wavelength coverage of either the HR13 or HR14B set-ups, especially for the likely members of NGC 1846.
Using the spectrum from the nearest sky fiber to a target did not work, nor did subtracting a combined spectrum
derived from all $16$ sky fibers in a given exposure. Using the {\sc iraf}
task {\sc skytweak} to vary the scaling of the sky spectra and apply small wavelength shifts improved the
results marginally, but not to a satisfactory level. The origin of the problem is unclear, but it may well
be linked to our inability to place sky fibers very locally to the NGC 1846 targets, as described in 
Section \ref{ss:targets}.

To work around this issue we used the {\sc iraf} task {\sc scombine} to merge all $96$ sky spectra for
a given set-up ($16$ sky fibers from each of $6$ exposures) into a high S/N ``master'' spectrum, and used 
this to generate a mask specifying all of the narrow wavelength intervals affected by sky lines. Across the
wavelength range of the HR13 set-up there were $33$ such intervals in the mask, covering $16.7\%$ of any
given spectrum, while for HR14B there were $25$ sky-line intervals excising $11.1\%$ of the coverage and for
HR11 there were $8$ intervals masking $3.3\%$. When detecting and measuring stellar absorption
features for our kinematic and chemical abundance analysis, we simply ignored any lines lying within one 
resolution element ($\sim 0.3\,$\AA) of a masked sky region. As described in Paper II, this procedure did 
not reject any lines critical for examining the abundance of a given element, but did ensure that any lines 
impacted by sky emission did not adversely influence our measurements.

\begin{figure*}
\begin{center}
\includegraphics[width=180mm]{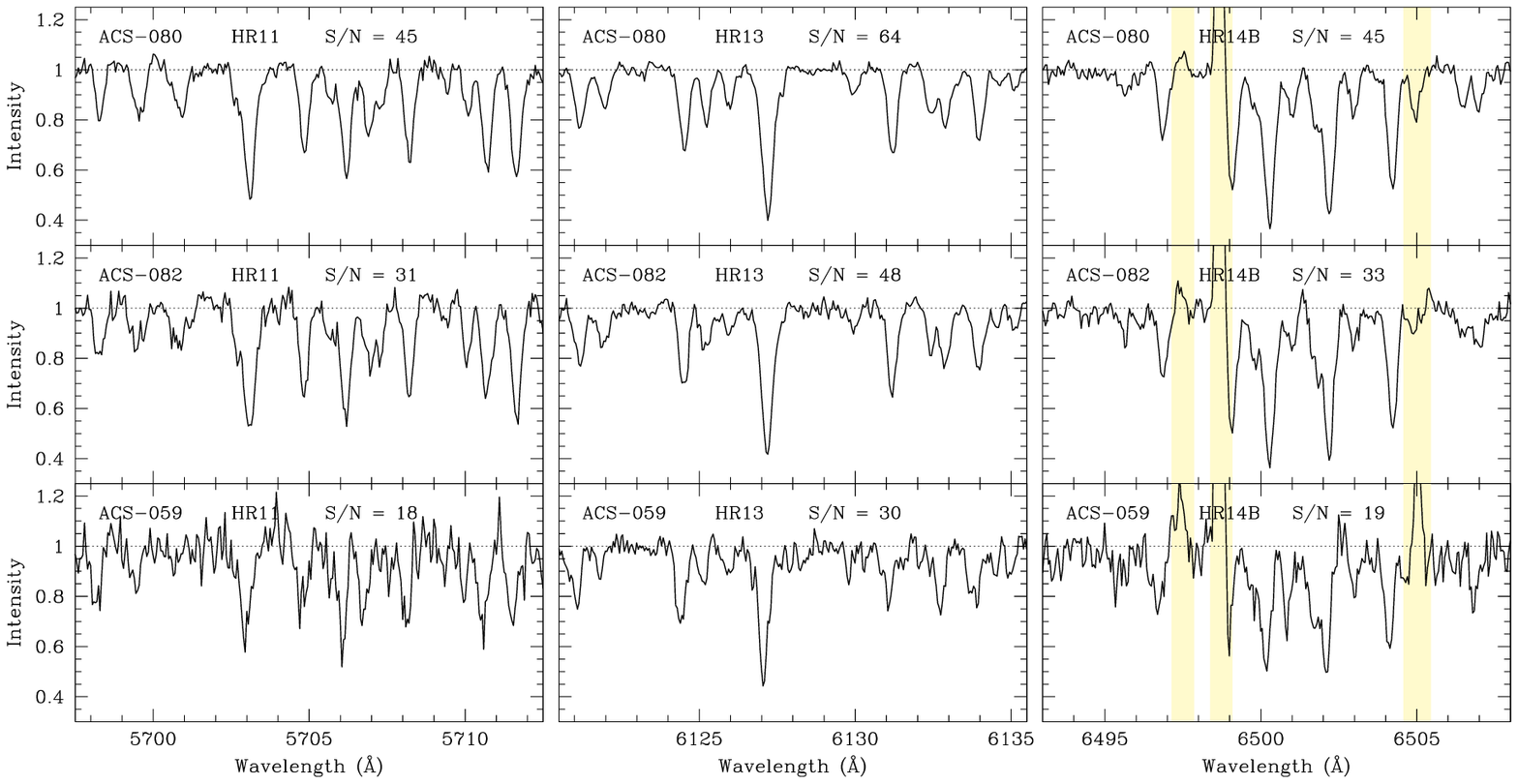} 
\end{center}
\caption{Representative spectral segments for three of our confirmed NGC 1846 members, as labelled. Each segment spans $\approx 15$\AA. The left column contains a region from the HR11 coverage, the central column from the HR13 coverage, and the right column from the HR14B coverage. The spectra have been normalised following our {\sc daospec} analysis (see Section \ref{ss:daospec}) and corrected to the heliocentric frame; however no correction has been made for the individual radial velocities. Regions masked in our analysis due to the presence of sky lines are shaded yellow (note that in this example such regions appear in the HR14B spectra only). The top row of the plot corresponds to one of the brightest targets in the NGC 1846 sample, and the spectra are of commensurately high S/N. Similarly, the bottom row shows spectra for one of the faintest targets, and these are of correspondingly low S/N. The middle row contains data of intermediate S/N; two-thirds of our sample of confirmed NGC 1846 members have spectra of comparable or higher S/N. Note that even in the lowest S/N spectra, most absorption lines are still clearly visible.\label{f:spectra}}
\end{figure*}

We encountered a similar problem when attempting to correct for telluric absorption in the HR13 set-up.
Four of our six bright blue targets turned out to be suitable stars for this type of correction; however
despite trying a similar variety of techniques as for the sky subtraction, we could not consistently 
achieve a high quality correction across the full affected wavelength range. Ultimately, we adpoted an identical 
approach to our sky-line solution -- we normalised then combined the $24$ telluric star exposures into a 
high S/N master spectrum and used this to generate a mask specifying the intervals affected by 
atmospheric absorption lines. There were $13$ such intervals covering $10.1\%$ of the HR13 wavelength
range; we note some overlap between the telluric and sky-line masks. Again, during our kinematic and 
chemical abundance analysis we simply ignored any stellar lines lying within one resolution element of a 
masked telluric-line region. As described in more detail in Paper II, the only critical stellar line rejected 
by this procedure was the $[$O{\sc i}$]$ line at $6300.3\,$\AA, which, for NGC 1846 targets, is redshifted 
onto a telluric feature near $6305.5\,$\AA. In order to use this line, we applied our master spectrum 
and the {\sc iraf} task {\sc telluric} to obtain a high-quality correction to only the $6305.5\,$\AA\ feature, 
and then used these specially tailored spectra when deriving oxygen abundances.

Finally, we corrected all on-target spectra to the heliocentric frame using the {\sc iraf} tasks {\sc rvcorrect}
and {\sc dopcor}. We then median-combined the six exposures of a given target in a given set-up using 
the {\sc iraf} task {\sc scombine}. Best results (including cosmic ray removal) were achieved
by scaling the input spectra to a common flux level and then, when combining them, weighting each by 
its median flux value and applying a sigma-clipping rejection algorithm. Wavelength regions covered by 
the sky- and telluric-line masks were excluded when computing the scaling and weighting factors.

Example spectra for three NGC 1846 stars spanning nearly the full range in brightness of our sample
of confirmed members (see Section \ref{ss:memstar}) are shown in Fig. \ref{f:spectra}. 
We estimated the continuum signal-to-noise (S/N) for these objects with the 
{\sc iraf} task {\sc splot}. The setting with the highest S/N per pixel is HR13; spectra for the HR11 and 
HR14B settings have S/N levels consistently $\sim 35-40\%$ lower for the same exposure duration. 
Partly this is due to the slightly higher resolution of these two settings (the HR11 and HR14B settings 
span wavelength ranges $\approx 15\%$ shorter than that for HR13 -- see Table \ref{t:flames}) and 
partly it is due to their somewhat lower overall efficiency. The HR13 spectra span S/N $\sim 30-70$
per pixel, while the HR11 and HR14B spectra have S/N in the range $\sim 18-50$ per pixel. For reference,
the HR13 setting has $\approx 5.6$ pixels per resolution element at the mid-point of its coverage, 
whereas HR11 has $\approx 4.7$ and HR14B $\approx 4.5$. Overall, two thirds of our confirmed NGC 1846
stars have spectra with S/N comparable to, or greater than, the middle star plotted in Fig. \ref{f:spectra}.

\subsection{Line Identification \& Measurement} 
\label{ss:daospec}
We next applied the {\sc daospec} software package to each of our $3\times(29+76) = 315$ GIRAFFE HR spectra.
{\sc daospec} is an automated tool, optimised for spectra with $R \ga 15\,000$, for identifying and measuring 
absorption lines. Full details may be found in \citet{stetson:08}, but briefly, it works by iteratively
finding lines in a given spectrum, fitting these with Gaussian profiles of fixed FWHM and subtracting them, 
and then using the residual spectrum to refine the continuum normalization and improve the line centroids
and strengths. Once this process has converged, the measured lines are cross-correlated against a 
user-supplied line list to provide an estimate of the radial velocity of the target along with an identification 
for any lines that are successfully matched. The accuracy of {\sc daospec} measurements, especially for
equivalent widths, has been tested and verified on GIRAFFE HR data \citep[e.g.,][]{pompeia:08,letarte:10}.

Full details of our equivalent width measurements are provided in Paper II; here we are mainly interested in
the radial velocities supplied by {\sc daospec}. We assembled an input line list and atomic data for
$18$ neutral and $12$ singly ionised species ($25$ elements in total) over the wavelength interval 
$5550-6650$\AA\ using the VALD atomic line database \citep{kupka:99}. We included only the ``strongest'' lines
for each species, where strength was approximately parametrized by the difference between the oscillator strength
$\log(gf)$ and the excitation potential $\chi$ (i.e., by the sum $\log(gf) - \chi$). For each species an
empirical minimum limit for this value was defined by manually examining lines in our highest S/N spectra 
and identifying those with the smallest equivalent widths that could be reliably identified and measured 
(typically $EW\approx 15$m\AA). Our final line list contained a total of $1445$ lines over the $1100$\AA\ 
wavelength interval. Before input to {\sc daospec} we applied our sky and telluric masks to cull affected lines 
from the list.

When measuring the radial velocity of a target, {\sc daospec} uses a sigma-clipping rejection algorithm
to eliminate identified lines with discrepant velocities. In order to determine precise radial velocities for
our stars we performed an initial run on all spectra with stringent $2\sigma$ rejection. This was important 
because at this stage we had not checked our line list for features that would be blended at the GIRAFFE 
HR resolutions. As described in Paper II, for the equivalent width measurements we subsequently performed 
a more relaxed $3.5\sigma$ rejection run and then determined the most appropriate lines for elemental 
abundance analysis by careful visual inspection of the spectra. However important additional guidance 
was provided by the identification of those lines initially rejected in the stringent $2\sigma$ {\sc daospec} run.

For each of our targets we ultimately obtained three independent radial velocity estimates from {\sc daospec}, 
one for each HR setting. Typically, $\approx 130$ lines were identified in each HR11 spectrum, $\approx 100$ lines 
in each HR13 spectrum and $70$ lines in each HR14B spectrum; however significant variation was seen as a function 
of S/N and the effective temperature of the target. In the majority of cases the line-by-line variance for each 
individual radial velocity estimate was in the range $0.5 \la \sigma \la 1.5$\ km$\,$s$^{-1}$; we conservatively adopted
this quantity to represent the uncertainty on each such measurement. 

While checking our 
results we noticed the presence of small systematic offsets between velocities measured for a given star from 
different settings. The $2\sigma$-clipped mean offsets were $V_{\rm HR13} - V_{\rm HR11} = 0.55$\ km$\,$s$^{-1}$ 
(97 stars) and $V_{\rm HR13} - V_{\rm HR14B} = -0.36$\ km$\,$s$^{-1}$ (94 stars). The origin of these offsets
is not clear, but they were straightforward to correct. For each setting we precisely measured the positions
of all the atmospheric emission features in each of the $16$ blank sky spectra. We then matched these
measurements against the wavelengths listed in the atlas of \citet{osterbrock:96} and derived the mean
($2.5\sigma$-clipped) offset. We found that the HR11 spectra needed to be shifted by $0.39$\ km$\,$s$^{-1}$,
the HR13 spectra by $-0.17$\ km$\,$s$^{-1}$, and the HR14B spectra by $-0.52$\ km$\,$s$^{-1}$.
No dependence on fiber number (i.e., the position of individual sky spectra on the FLAMES CCD) was evident. 
Applying these corrections almost completely removed the mean systematic offsets between velocities 
measured from the three different settings for a given target. We combined these corrected quantities in a 
weighted average to obtain a final radial velocity for each star, along with its associated uncertainty.

\begin{figure}
\begin{center}
\includegraphics[width=86mm]{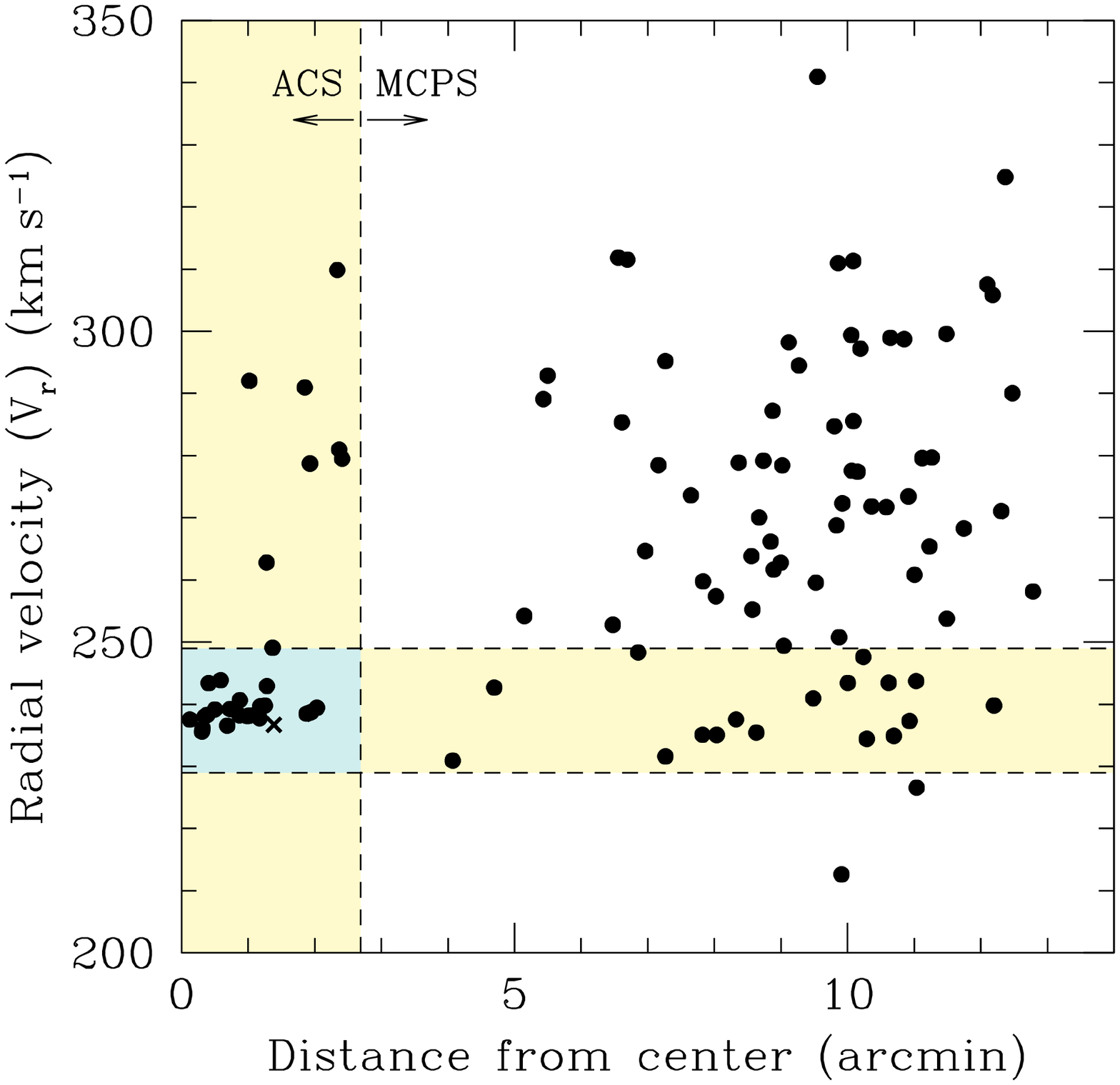} \\
\vspace{2mm}
\includegraphics[width=86mm]{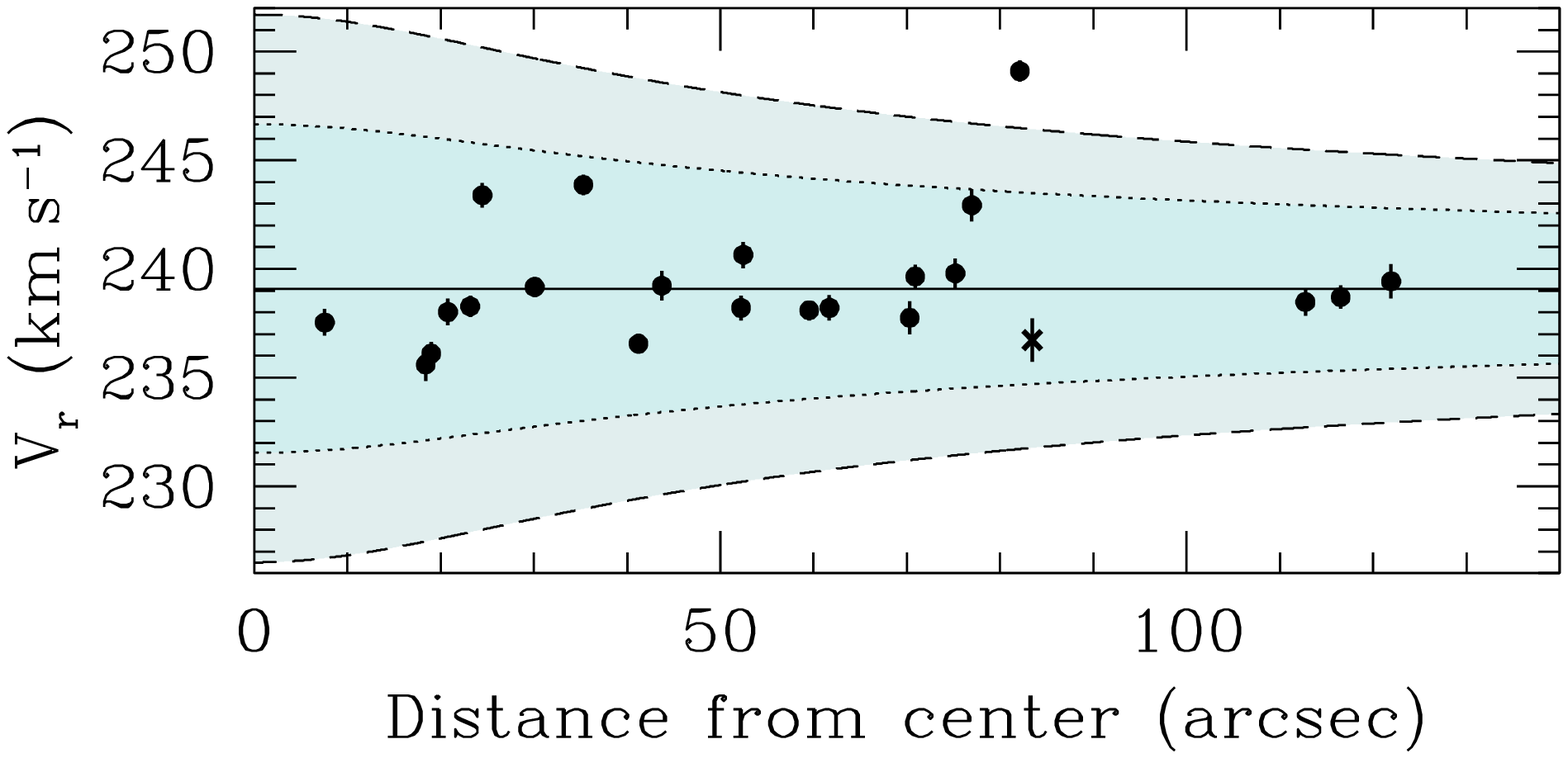}
\end{center}
\caption{Radial velocity versus distance from the center of NGC 1846 for all of our FLAMES science targets. The upper panel spans the full radial coverage, and the full range of radial velocities. The vertical shaded area indicates the region lying inside the nominal boundary of the cluster, while the horizontal shaded region denotes our initial radial velocity cut of $\pm 10$\ km$\,$s$^{-1}$ about the mean velocity of targets lying within $r_h = 34.5\arcsec$. The lower panel zooms in on the intersection of these two regions. Here, the shaded areas indicate $\pm 5$ and $\pm 3$ times the velocity dispersion at given radius, calculated according to Eq. \ref{e:plummer} (see also Figure \ref{f:dispersion}). In both panels the position of the planetary nebula Mo-17 is marked with a cross.\label{f:members}}
\end{figure}
 
\begin{deluxetable*}{cccccccccccccc}
\tabletypesize{\scriptsize}
\tablecaption{Data on the radial velocity members of NGC 1846 studied in this work.\label{t:members}}
\tablehead{
\colhead{Target} & \hspace{2mm} & \multicolumn{2}{c}{Position (J2000.0)} & \colhead{Radius} & \colhead{PA} & \hspace{2mm} & \colhead{$B$} & \colhead{$V$} & \colhead{$I$} & \hspace{2mm} & \colhead{Velocity} & \hspace{2mm} & \colhead{$P_{mem}$}\\
\colhead{Name} & & \colhead{RA} & \colhead{Dec} & \colhead{$(\arcsec)$} & \colhead{$(\degr)$} & & \colhead{} & \colhead{} & \colhead{} & & \colhead{(km$\,$s$^{-1}$)} & &
}
\startdata
ACS-001 & & $05\;07\;36.8$ & $-67\;27\;45.9$ & $19.0$  & $105.8$ & & $18.16$ & $16.56$ & $14.86$ & & $236.1 \pm 0.5$ & & $0.996$ \\
ACS-013 & & $05\;07\;33.6$ & $-67\;26\;41.2$ & $59.5$  & $359.6$ & & $18.35$ & $16.99$ & $15.56$ & & $238.1 \pm 0.5$ & & $0.995$ \\
ACS-017 & & $05\;07\;38.4$ & $-67\;28\;11.7$ & $41.2$  & $138.7$ & & $18.51$ & $17.18$ & $15.73$ & & $236.6 \pm 0.4$ & & $0.990$ \\
ACS-025 & & $05\;07\;36.2$ & $-67\;27\;58.8$ & $23.2$  & $141.2$ & & $18.63$ & $17.28$ & $15.93$ & & $238.3 \pm 0.5$ & & $0.999$ \\
ACS-030 & & $05\;07\;39.0$ & $-67\;28\;23.2$ & $52.5$  & $144.1$ & & $18.66$ & $17.43$ & $16.12$ & & $240.7 \pm 0.6$ & & $0.985$ \\
ACS-036 & & $05\;07\;30.4$ & $-67\;29\;35.7$ & $116.5$ & $189.4$ & & $18.86$ & $17.63$ & $16.38$ & & $238.7 \pm 0.5$ & & $0.988$ \\
ACS-043 & & $05\;07\;21.6$ & $-67\;27\;25.3$ & $70.9$  & $282.5$ & & $18.84$ & $17.70$ & $16.45$ & & $239.7 \pm 0.5$ & & $0.994$ \\
ACS-046 & & $05\;07\;32.6$ & $-67\;27\;45.5$ & $7.6$   & $230.5$ & & $19.11$ & $17.92$ & $16.68$ & & $237.5 \pm 0.6$ & & $0.997$ \\
ACS-047 & & $05\;07\;34.5$ & $-67\;28\;24.2$ & $43.7$  & $173.7$ & & $19.02$ & $17.84$ & $16.62$ & & $239.2 \pm 0.7$ & & $0.997$ \\
ACS-051 & & $05\;07\;36.6$ & $-67\;27\;33.5$ & $18.4$  & $66.8$  & & $19.10$ & $17.96$ & $16.71$ & & $235.6 \pm 0.8$ & & $0.992$ \\
ACS-053 & & $05\;07\;29.6$ & $-67\;26\;27.4$ & $77.0$  & $342.3$ & & $19.09$ & $17.89$ & $16.65$ & & $242.9 \pm 0.8$ & & $0.919$ \\
ACS-059 & & $05\;07\;28.6$ & $-67\;28\;44.8$ & $70.3$  & $204.3$ & & $19.15$ & $18.04$ & $16.85$ & & $237.8 \pm 0.8$ & & $0.975$ \\
ACS-066 & & $05\;07\;49.6$ & $-67\;29\;01.4$ & $121.9$ & $131.5$ & & $...$   & $18.18$ & $17.00$ & & $239.4 \pm 0.8$ & & $0.988$ \\
ACS-080 & & $05\;07\;30.3$ & $-67\;27\;11.2$ & $35.3$  & $326.8$ & & $18.32$ & $17.09$ & $15.73$ & & $243.9 \pm 0.5$ & & $0.974$ \\
ACS-081 & & $05\;07\;30.4$ & $-67\;28\;04.2$ & $30.1$  & $218.8$ & & $18.45$ & $17.10$ & $15.74$ & & $239.2 \pm 0.4$ & & $0.998$ \\
ACS-082 & & $05\;07\;30.1$ & $-67\;27\;27.4$ & $24.5$  & $303.1$ & & $18.47$ & $17.14$ & $15.81$ & & $243.4 \pm 0.6$ & & $0.984$ \\
ACS-085 & & $05\;07\;30.0$ & $-67\;26\;42.7$ & $61.7$  & $340.1$ & & $18.53$ & $17.15$ & $15.81$ & & $238.2 \pm 0.6$ & & $0.995$ \\
ACS-090 & & $05\;07\;30.2$ & $-67\;27\;46.1$ & $20.8$  & $254.9$ & & $18.76$ & $17.53$ & $16.29$ & & $238.0 \pm 0.6$ & & $0.998$ \\
ACS-092 & & $05\;07\;28.7$ & $-67\;28\;24.4$ & $52.2$  & $213.3$ & & $18.75$ & $17.57$ & $16.33$ & & $238.2 \pm 0.6$ & & $0.996$ \\
ACS-102 & & $05\;07\;27.4$ & $-67\;29\;27.6$ & $112.7$ & $198.6$ & & $18.97$ & $17.90$ & $16.75$ & & $238.5 \pm 0.6$ & & $0.988$ \\
ACS-112 & & $05\;07\;43.3$ & $-67\;26\;49.9$ & $75.2$  & $47.5$  & & $...$   & $18.02$ & $16.90$ & & $239.8 \pm 0.7$ & & $0.993$ \\
Mo-17    & & $05\;07\;25.3$ & $-67\;28\;51.0$ & $83.5$  & $214.3$ & & $...$   & $...$   & $...$   & & $236.7 \pm 1.5$ & & $0.967$
\enddata
\end{deluxetable*}

\begin{deluxetable}{ccccc}
\tabletypesize{\scriptsize}
\tablecaption{Data on non-members of NGC 1846.\label{t:nonmembers}}
\tablehead{
\colhead{Target} & \multicolumn{2}{c}{Position (J2000.0)} & \colhead{Radius} & \colhead{Velocity}\\
\colhead{Name} & \colhead{RA} & \colhead{Dec} & \colhead{$(\arcsec)$} & \colhead{(km$\,$s$^{-1}$)}
}
\startdata
ACS-019 & $05\;07\;14.5$ & $-67\;28\;16.4$ & $115.8$ & $278.7 \pm 0.5$ \\
ACS-022 & $05\;07\;20.0$ & $-67\;29\;36.9$ & $140.4$ & $309.9 \pm 0.7$ \\
ACS-024 & $05\;07\;21.1$ & $-67\;27\;01.3$ & $82.1$ & $249.1 \pm 0.5$ \\
ACS-026 & $05\;07\;23.6$ & $-67\;26\;50.6$ & $76.6$ & $262.8 \pm 0.5$ \\
ACS-029 & $05\;07\;14.7$ & $-67\;29\;15.8$ & $144.8$ & $279.5 \pm 0.5$ \\
ACS-054 & $05\;07\;26.3$ & $-67\;26\;56.8$ & $61.2$ & $292.0 \pm 0.8$ \\
ACS-070 & $05\;07\;10.5$ & $-67\;28\;31.1$ & $142.1$ & $281.0 \pm 0.8$ \\
ACS-072 & $05\;07\;14.4$ & $-67\;27\;48.1$ & $111.1$ & $291.0 \pm 0.9$ 
\enddata
\end{deluxetable}

\section{Cluster Membership}
\label{s:member}
\subsection{Stars within the truncation radius}
\label{ss:memstar}
The upper panel of Figure \ref{f:members} shows radial velocity versus projected distance 
from the center of NGC 1846 for all $29$ ACS stars and $76$ MCPS stars. %, and the planetary nebula Mo-17.
We adopted $\alpha = 05$:$07$:$33.66$, $\delta = -67$:$27$:$40.7$ for the cluster center, determined using 
our ACS photometry in the FK5 astrometric frame.
The tight grouping of objects to the lower left of the plot is indicative of the cluster. To determine membership
we imposed a maximum allowed radius of $161\arcsec$, corresponding to the truncation radius measured 
for NGC 1846 by \citet{goudfrooij:09}. We also calculated the mean velocity $\bar{V}$ of targets lying within the
half-light radius of $r_h = 34.5\arcsec$ obtained by the same authors, on the basis that these objects
are the most likely to be cluster members, and then imposed a generous radial velocity cut of 
$\pm 10$ km$\,$s$^{-1}$ about this value\footnote{We note that according 
to the 2010 update of the \citet{harris:96} online catalogue, all but the most massive Galactic globular clusters 
have central velocity dispersions of only a few km$\,$s$^{-1}$ \citep[see also][]{lane:09,lane:10a,lane:10b}.}. 
These criteria resulted in the exclusion of 
seven ACS stars, with one additional star lying on the boundary of the allowed region of parameter space.
The lower panel of Figure \ref{f:members} shows a close-up of this region. Although the candidate
star has a radial velocity only $\approx 10$ km$\,$s$^{-1}$ higher than the systemic velocity
of NGC 1846, this corresponds to more than five times the velocity dispersion at this radius within
the cluster (see Section \ref{s:kinematics} below) and we therefore also excluded this object.

Our $21$ probable stellar members of NGC 1846 are listed in Table \ref{t:members} along with 
their ACS/WFC photometry and measured radial velocities. Note, from the upper panel of 
Figure \ref{f:members}, that the velocity of the cluster overlaps significantly with the range
measured for non-members (i.e., those objects outside the truncation radius). Thus, we cannot be
certain that our assumed sample is entirely free of field interlopers. To assess the likelihood
of this, we calculated an indicative membership probability for all $21$ stars in our sample.
For each object, we used the cluster surface density profile of \citet{goudfrooij:09} (see the 
beginning of Section \ref{s:kinematics}, below) to estimate the likelihood that the star under 
consideration could be a non-member based on the relative densities of the cluster and field
at the appropriate radius. We then counted how many of the $76$ MCPS stars from outside
$r_t$ lay within the interval $\bar{V} \pm |v_{i} + \sigma_{i}|$, where $v_i$ and $\sigma_i$
are the velocity and associated uncertainty for the star under consideration, and used this
information to estimate the likelihood that the star could be a non-member based on the
deviation of its velocity from the cluster mean. We defined the final membership probability 
for the star, $P_{mem}$, to be the complement of the product of this quantity and that estimated 
from the density profile.

The values of $P_{mem}$ are listed in Table \ref{t:members}. Note that these are indicative
lower limits only. They are subject to stochastic fluctuations due to the small number
of MCPS objects lying comparably near to the cluster mean velocity, and do not include any
information about the photometric selection criteria we employed in defining our initial
sample of targets (Figure \ref{f:cmds}) -- which would effectively reduce the number of viable 
MCPS stars within the allowed velocity intervals by a factor $\ga 2$.

The membership probabilities are very high for all objects. The most likely non-member is 
ACS-053, which has a membership probability of $\approx 92\%$; for most others, 
$P_{mem} \ga 99\%$. These simple calculations give us confidence that we have defined a clean
set of NGC 1846 members.
We consider the remainder of our ACS and MCPS targets to belong to the LMC field (or, possibly, the 
Galactic foreground); further study of these objects is beyond the scope of the present work. 

In order to aid future observers of NGC 1846 
we list, in Table \ref{t:nonmembers}, those stars lying within the truncation radius of $161\arcsec$ 
which are not radial velocity members of the cluster. Members and non-members are also distinguished
on the CMDs in Figure \ref{f:cmds}.

\subsection{The Planetary Nebula Mo-17}
\label{ss:pn}
We briefly consider the planetary nebula Mo-17 in more detail. This object was first catalogued by 
\citet{morgan:94}, and, based on its proximity to
NGC 1846 (it lies at a radius of $83.5\arcsec \approx 20.3$\ pc), \citet{kontizas:96} suggested 
that it might in fact belong to the cluster. The catalogue of \citet{reid:06} lists a radial velocity of 
$241.8\pm 20.0$\ km$\,$s$^{-1}$ which is consistent with literature estimates for the motion 
of NGC 1846. %: $240\pm 10$\ km$\,$s$^{-1}$ \citep{olszewski:91}, and $235.2\pm 0.9$\ km$\,$s$^{-1}$ 
%\citep{grocholski:06}. 

\begin{figure}
\begin{center}
\includegraphics[width=86mm]{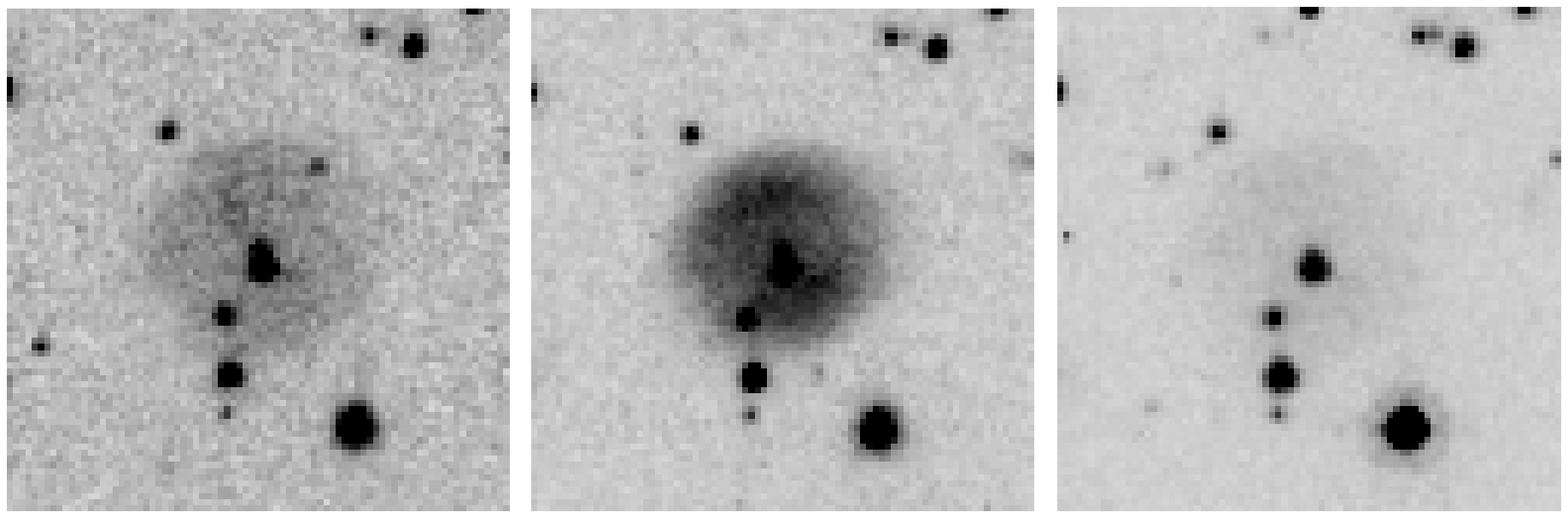} 
\end{center}
\caption{HST ACS/WFC images of the NGC 1846 planetary nebula Mo-17 -- from left to right, taken with the F435W, F555W, and F814W filters. Each thumbnail is $4\arcsec$ on a side.\label{f:pnimages}}
\end{figure}
 
Figure \ref{f:pnimages} shows ACS/WFC thumbnails of Mo-17 in the F435W, F555W and F814W 
passbands\footnote{Note that \citet{shaw:06} have previously reported on archival HST/ACS imaging
of this object; however based on the coordinates and image they present, a nearby face-on background
spiral galaxy was apparently misidentified as the planetary nebula. It is not clear whether their brief 
description of the appearance of Mo-17 corresponds to the correct object or this interloper.}.
The nebula is circular in appearance, with a diameter of approximately 
$1.3\arcsec \approx 0.3$\ pc. It is brightest in the F555W image, in which it also appears to have
two distinct lobes, or perhaps an outer ring. The central star may be visible -- there is a blend of two 
objects lying near the middle of the nebula. The fainter of these (the upper object in Fig. \ref{f:pnimages})
is much bluer than the brighter; however due to their close proximity, precise photometry is 
very difficult. 

\begin{figure}
\begin{center}
\includegraphics[width=86mm]{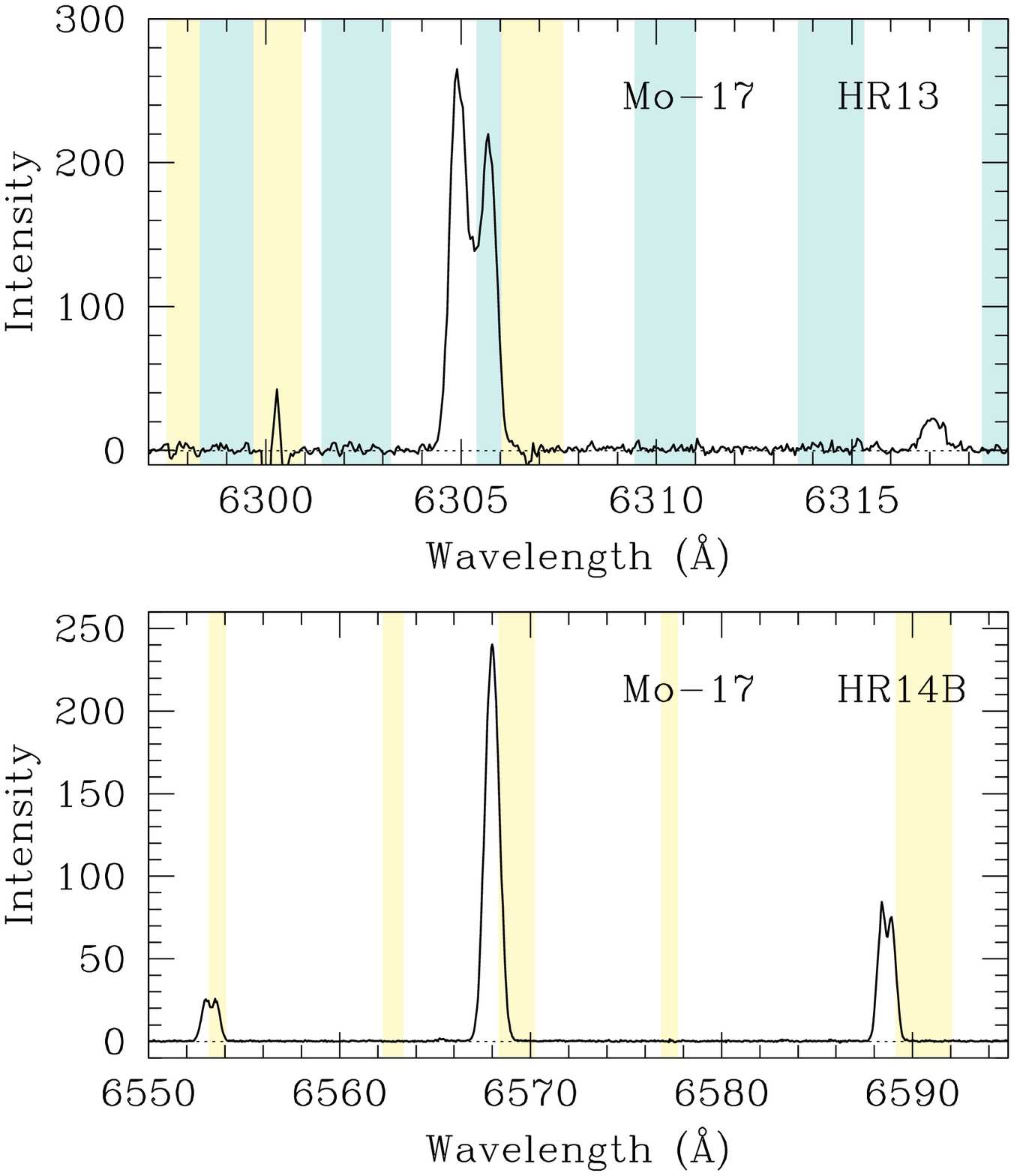} 
\end{center}
\caption{Example HR13 and HR14B spectral segments for the NGC 1846 planetary nebula Mo-17. As in Fig. \ref{f:spectra} these have been corrected to the heliocentric frame, but no correction for the radial velocity of the target has been made. Note that the intensity scale is in arbitrary units. We have subtracted the sky lines as accurately as possible across these small segments. As before, regions in the sky line mask are shaded yellow. Note that most lines have subtracted cleanly; however strong residuals are present for the $[$O{\sc i}$]$ line near $6300$\AA. Regions masked due to the presence of telluric absorption are shaded blue. The HR13 panel spans $\sim 22$\AA; nebular $[$O{\sc i}$]$ emission is visible at $6305$\AA; note that this line is broadened and double-peaked. Weak $[$S{\sc iii}$]$ emission is also visible near $6317$\AA. The HR14B panel spans a larger spectral range of $\sim45$\AA. Double-peaked $[$N{\sc ii}$]$ emission lines are visible at either end of the segment, with a strong single-peaked H$\alpha$ line near $6568$\AA. This line was used to derive the velocity of Mo-17.\label{f:pnspec}}
\end{figure}

As noted previously, we allocated a MEDUSA fiber to Mo-17. The fiber diameter of $1.2\arcsec$ corresponds
well to the size of the nebula. Example spectra from the HR13 and HR14B
settings are plotted in Fig. \ref{f:pnspec}. Although there is sparse contamination from sky line residuals,
in the HR13 setting $[$O{\sc i}$]\,6300$\AA\ emission is clearly visible, while in the HR14B spectrum
there is strong emission from the $[$N{\sc ii}$]\,6548$\AA\ and $6583$\AA\ lines and H$\alpha$.
Also detected in the HR13 spectrum is a weak $[$S{\sc iii}$]\,6312$\AA\ line, and the 
$[$O{\sc i}$]\,6363$\AA\ line (not plotted). The HR11 setting (not shown) covers only one weak line,
$[$N{\sc ii}$]\,5755$\AA, while 
the LR02 spectrum (also not shown) exhibits a variety of emission features including the Balmer lines
H$\gamma$, H$\delta$ and H$\epsilon$, several He{\sc i} lines, $[$O{\sc iii}$]\,4363$\AA,
and $[$Ne{\sc iii}$]\,3967$\AA. %We plan to report on these spectra in more detail in a future work.

In Fig. \ref{f:pnspec} is it quite evident that the strong $[$O{\sc i}$]\,6300$\AA\ and $[$N{\sc ii}$]\,6548$\AA, 
$6583$\AA\ lines are broad and double-peaked. The $[$O{\sc i}$]\,6363$\AA\ line also follows
this pattern. It is likely that this reflects the lobed structure seen in the F555W image in Fig. \ref{f:pnimages}.
The F555W filter spans the range $\sim 4700-6100$\AA, within which emission from 
$[$O{\sc iii}$]\,4959$\AA\ and $5007$\AA\ will be dominant. The separation of the peaks in the two 
$[$O{\sc i}$]$ lines indicates outflow velocities of approximately $\pm 20$\ km$\,$s$^{-1}$, while the 
$[$N{\sc ii}$]$ peaks suggest $\pm 10$\ km$\,$s$^{-1}$. 

The velocity structure present in these emission lines makes it difficult to obtain a precise estimate
for the radial velocity of Mo-17. Notably however, the H$\alpha$ line, while equivalently broad as the
lines discussed above, has only a single peak and is highly symmetric. H$\alpha$\ falls between the
wavelength coverage of the ACS/WFC F555W and F814W filters. However, light in the F435W image of the 
nebula appears much more evenly distributed than in F555W; we know from our LR02 spectra that emission 
in the range covered by this filter ($\sim 3700-4800$\AA) comes predominantly from the Balmer series.
Given this, and the circular symmetry of the nebula, we make the assumption that the peak of the
H$\alpha$ line reflects the overall radial motion of Mo-17, and derive a heliocentric velocity
$V_r = 236.7$\ km$\,$s$^{-1}$. Estimating the mid-points of the two $[$O{\sc i}$]$ and two
$[$N{\sc ii}$]$ lines and taking the average of these measurements leads to a velocity 
within $\approx 1.2$\ km$\,$s$^{-1}$ of this, and we thus adopt an overall uncertainty of 
$\pm 1.5$\ km$\,$s$^{-1}$ on our result. As plotted in Figure \ref{f:members}, Mo-17 is very likely a 
member of NGC 1846; the indicative membership probability, $P_{mem} \approx 97\%$. 
We include this object in Table \ref{t:members}.

\section{Cluster Kinematics}
\label{s:kinematics}
Our radial velocity measurements allowed us to investigate the internal dynamics of NGC 1846.
In what follows we make use of the cluster structural parameters measured by \citet{goudfrooij:09}. 
These authors fit a \citet{king:62} model:
\begin{equation}
n(r_p) = n_0 \left[ \frac{1}{\sqrt{1 + (r_p/r_c)^2}} - \frac{1}{\sqrt{1 + (r_t/r_c)^2}} \right]^2
\label{e:king}
\end{equation} 
to a radial number-density profile constructed from 
their ACS/WFC imaging of the cluster. They found a  core radius $r_c = 26.0\arcsec$ and a truncation radius 
$r_t = 161.2\arcsec$. They also measured $r_h = 34.5\arcsec$ -- assuming the cluster is not strongly 
segregated by luminosity (or mass), this is a good estimate of its projected half-light (or half-mass) radius.
Adopting the canonical LMC distance modulus $\mu = 18.5$, these values correspond to $r_c = 6.3$\ pc, 
$r_h = 8.4$\ pc, and $r_t = 39.2$\ pc. %\citet{goudfrooij:09} also measured the central stellar density
%of the cluster to be $n(0) = 5.73$\ arcsec$^{-2}$, along with a local field star density (assumed to be
%uniform across the ACS/WFC field of view) of $n_f = 0.27$\ arcsec$^{-2}$.

\subsection{Rotation}
\label{ss:rotation}
We first checked for any signature of rotation within the cluster, following a commonly used procedure 
\citep[see e.g.,][and references therein]{lane:09,lane:10a,lane:10b,bellazzini:12}. We calculated the
position angle (PA, east of north) of each member in the plane of the sky with respect to the cluster center, 
then split the sample with the dividing line PA$\;=0\degr-180\degr$ and calculated the difference in
mean radial velocity between the two sub-groups of stars ($\Delta\bar{V}_r$). This process was then
repeated with the position angle of the dividing line incremented by $20\degr$ each time. 

If coherent rotation is present in the cluster and does not lie entirely in the plane of the sky, this should 
manifest in the form of a sinusoidal pattern when $\Delta\bar{V}_r$ is plotted as a function of the position 
angle of the dividing line. The position angle at which the maximum amplitude of this sinusoid occurs 
corresponds to the projected axis of rotation, while the amplitude itself represents twice that of the mean 
rotation -- i.e., $A_{\rm rot} = \Delta\bar{V}_r / 2$. The observed (projected) amplitude $A_{\rm rot}$ is a lower 
limit to the true amplitude of rotation $A_{\rm true}$ as there is a correction factor $\sin i$ to consider, where 
$i$ is the inclination of the rotation axis with respect to the plane of the sky ($i=90\degr$ represents an 
``edge-on'' cluster, where the observed amplitude of rotation would match the true amplitude, while 
$i=0\degr$ is a pole-on cluster where no rotation would be seen from radial velocity measurements 
irrespective of the true amplitude). 
%As noted by \citet{bellazzini:12}, because the sine function maps a 
%uniform distribution of $i$ into a distribution that is strongly peaked towards $\sin i = 1$, the observed 
%rotation amplitude is, statistically speaking, a reasonable proxy for the true amplitude.

\begin{figure}
\begin{center}
\includegraphics[width=86mm]{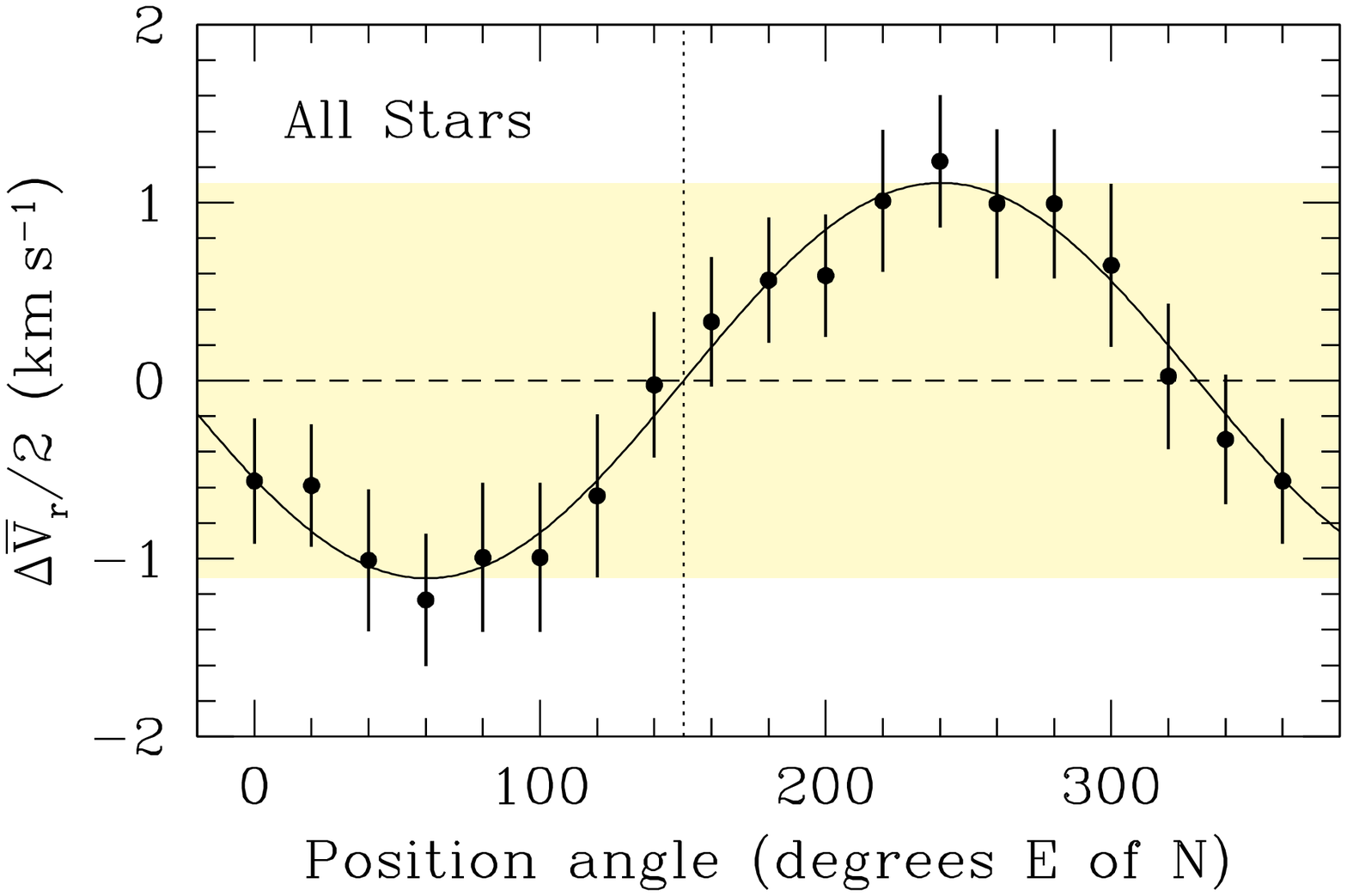} \\
\vspace{2mm}
\includegraphics[width=86mm]{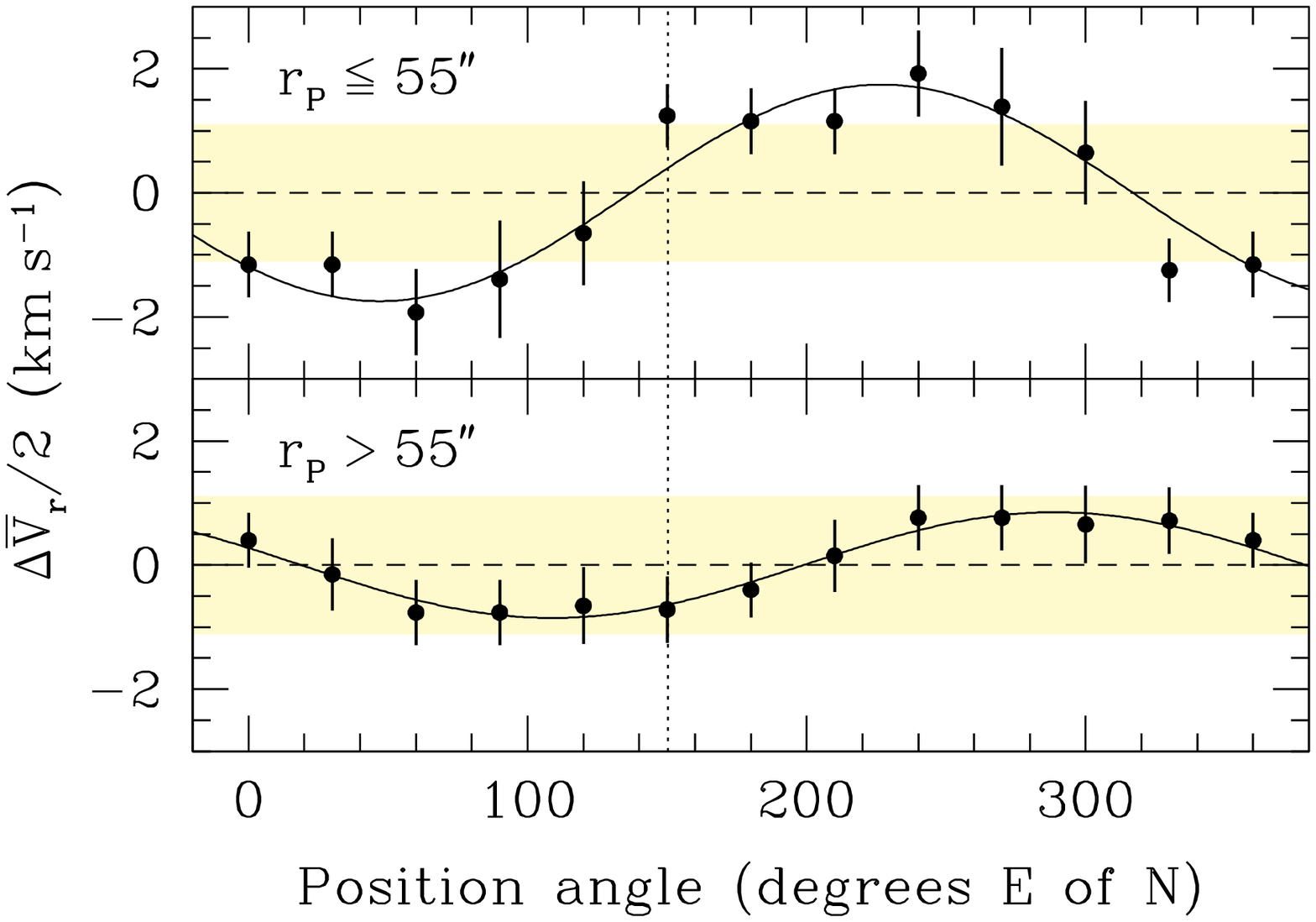}
\end{center}
\caption{Plots of $\Delta\bar{V}_r / 2$ as a function of the position angle of the dividing line for stars in NGC 1846. The upper panel shows the measurements and best-fitting model for the entire sample -- note the clean sinusoidal shape of the rotation curve. The shaded region indicates the measured amplitude of rotation,  $A_{\rm rot} = 1.1$\ km$\,$s$^{-1}$. In the lower two panels we divide the sample in half using a radius $r_p = 55\arcsec$ and repeat the measurements. The inner half of the sample appears to rotate much more strongly than the outer half. For comparison to the upper panel, the shaded region again denotes the amplitude of rotation measured for the full sample.\label{f:rotate}}
\end{figure}
 
The upper panel of Figure \ref{f:rotate} shows the results for our sample of $22$ tracers in NGC 1846.
We plot $\Delta\bar{V}_r / 2$ in order to more clearly elucidate the observed rotation amplitude.
To determine the point-to-point uncertainties, we used a bootstrapping method whereby we generated
a large number of mock systems using our observed sample. Each mock system consisted of $22$
stellar velocities chosen randomly from our $22$ measurements, with repeated selection allowed.
Every time we selected a velocity $v_i$ we also generated a random Gaussian deviate of $v_i$ (that is,
a value from a Gaussian distribution with $\sigma_{i}$ -- i.e., the measurement uncertainty in $v_i$) 
and added this to the velocity. We then
repeated our measurement of $\Delta\bar{V}_r$ as a function of position angle for all mock systems,
with the distribution of measurements at given PA indicative of the uncertainty at that point. This procedure
allowed us to account naturally for both the (relatively small) size of our sample and the individual
uncertainties on our measured radial velocities.

Our results show a clear sinusoidal pattern, suggestive that systemic rotation is present in NGC 1846. 
The weighted best fit has $A_{\rm rot} = 1.1$\ km$\,$s$^{-1}$ and the axis of rotation lying along the line
PA$\;=60\degr-240\degr$. Uncertainties derived directly from this fit likely under-estimate the true
uncertainties, because the points are all correlated. \citet{bellazzini:12} adopt $\pm 0.5$\ km$\,$s$^{-1}$ 
and $\pm 30\degr$ as conservative $1\sigma$ uncertainties on their measured rotation amplitudes and axis 
orientations, based on experimentation with sub-samples of their observed stars and comparisons with external 
samples. We are not in a position to conduct similar tests; instead we investigated this problem using another 
large set of mock systems. 

The question is how strongly we can constrain the amplitude and orientation of rotation in the cluster given 
the sample of stars we have measured. To this end each mock system consisted of $22$ stars at the same 
radius and PA as in the real sample. We assumed simple cylindrical rotation with amplitude and orientation as
measured above -- this defined a velocity for each star based on its PA, to which we added a random 
measurement uncertainty as previously, and a second random component defined by the velocity dispersion profile
of NGC 1846 (see Section \ref{ss:meandisp}, below). We then followed the same procedure outlined above
to measure the rotation amplitude and orientation, and examined the distributions of these recovered
quantities over all mock systems to determine our uncertainties. We found the measured amplitude of
rotation to be uncertain at the $\pm 0.4$\ km$\,$s$^{-1}$ level, and the orientation of the rotation axis
to be uncertain at the $\pm 20\degr$ level -- similar to the values adopted by \citet{bellazzini:12}.

We next decided to test whether there was any change in the degree of rotation with radius within the 
cluster. To do this we split our sample in half using a dividing radius $r_p = 55\arcsec$, which corresponds
to roughly $1.5$ times the half-light radius, and repeated our measurements
using only stars within, and then outwith, this radius. The results are shown in the lower panel of Figure
\ref{f:rotate}. Although the samples are by now rather small, there is an indication that the rotation
amplitude for the inner sample is significantly larger than that for the outer sample -- the best fits
have amplitudes of $1.8 \pm 0.6$\ km$\,$s$^{-1}$ and $0.8 \pm 0.4$\ km$\,$s$^{-1}$ respectively.
The PA of the rotation axis changes by only a small amount for the inner sample, to $47 \pm 25\degr$,
but increases substantially for the outer sample, to $109 \pm 35\degr$. The quoted uncertainties
were again derived using large sets of mock systems.

\citet{bellazzini:12} note that variation of rotation amplitude with radius is not unusual in Galactic 
globular clusters, such that the rotation amplitude determined using stars at all radii typically under-estimates 
the {\it maximum} rotation amplitude in the cluster. Their experiments indicate that $\Delta\bar{V}_r = 2A_{\rm rot}$,
where $A_{\rm rot}$ is calculated (as we have done here) using stars at all radii,
is likely to be a better estimate of the maximum rotation amplitude. This is consistent with the lower 
panel of Figure \ref{f:rotate}, which suggests that the maximum projected rotation amplitude for NGC 1846 could
be as high as $\approx 2$\ km$\,$s$^{-1}$.

\begin{figure}
\begin{center}
\includegraphics[width=83mm]{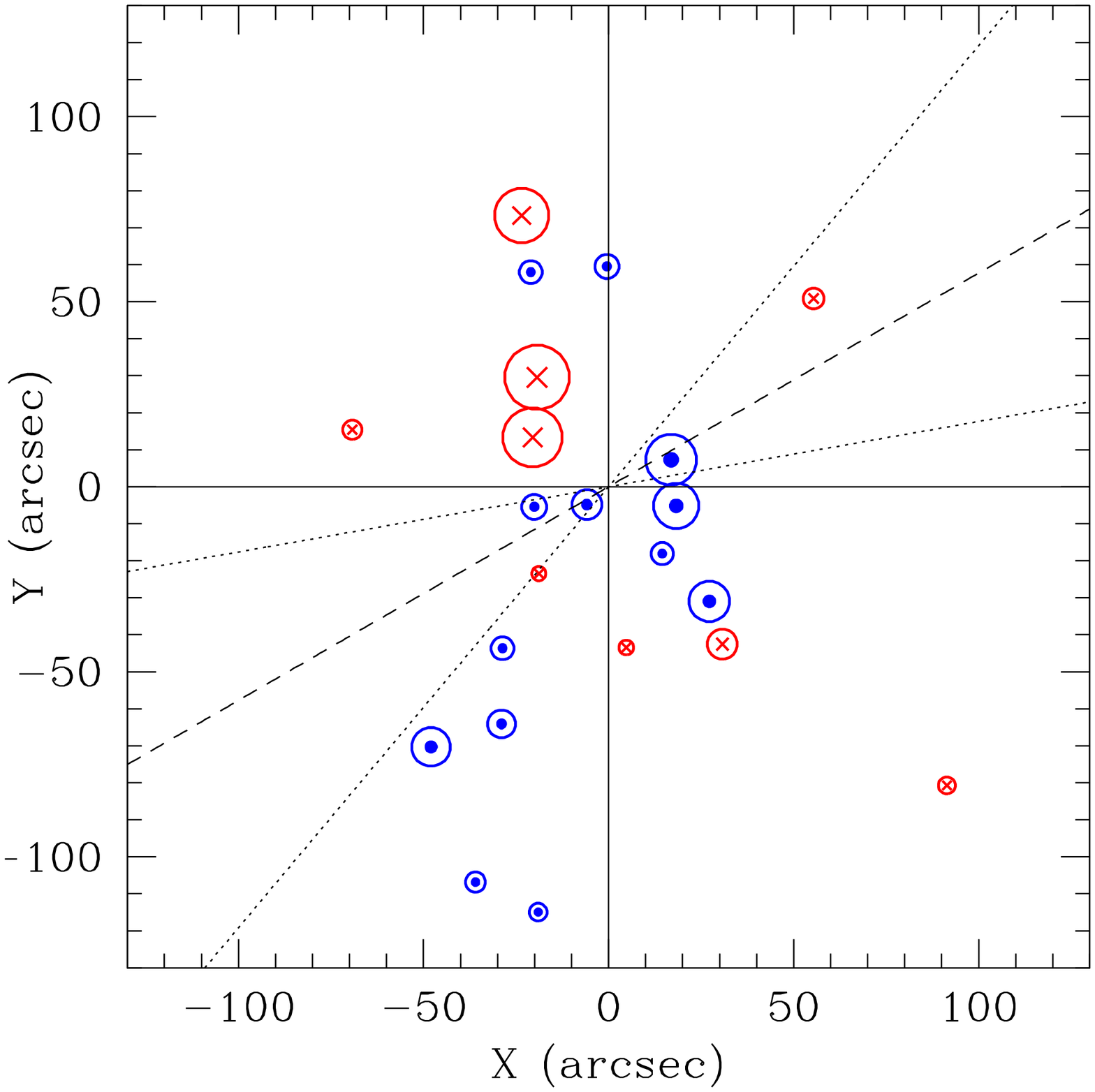} \\
\vspace{-5mm}
\includegraphics[width=83mm]{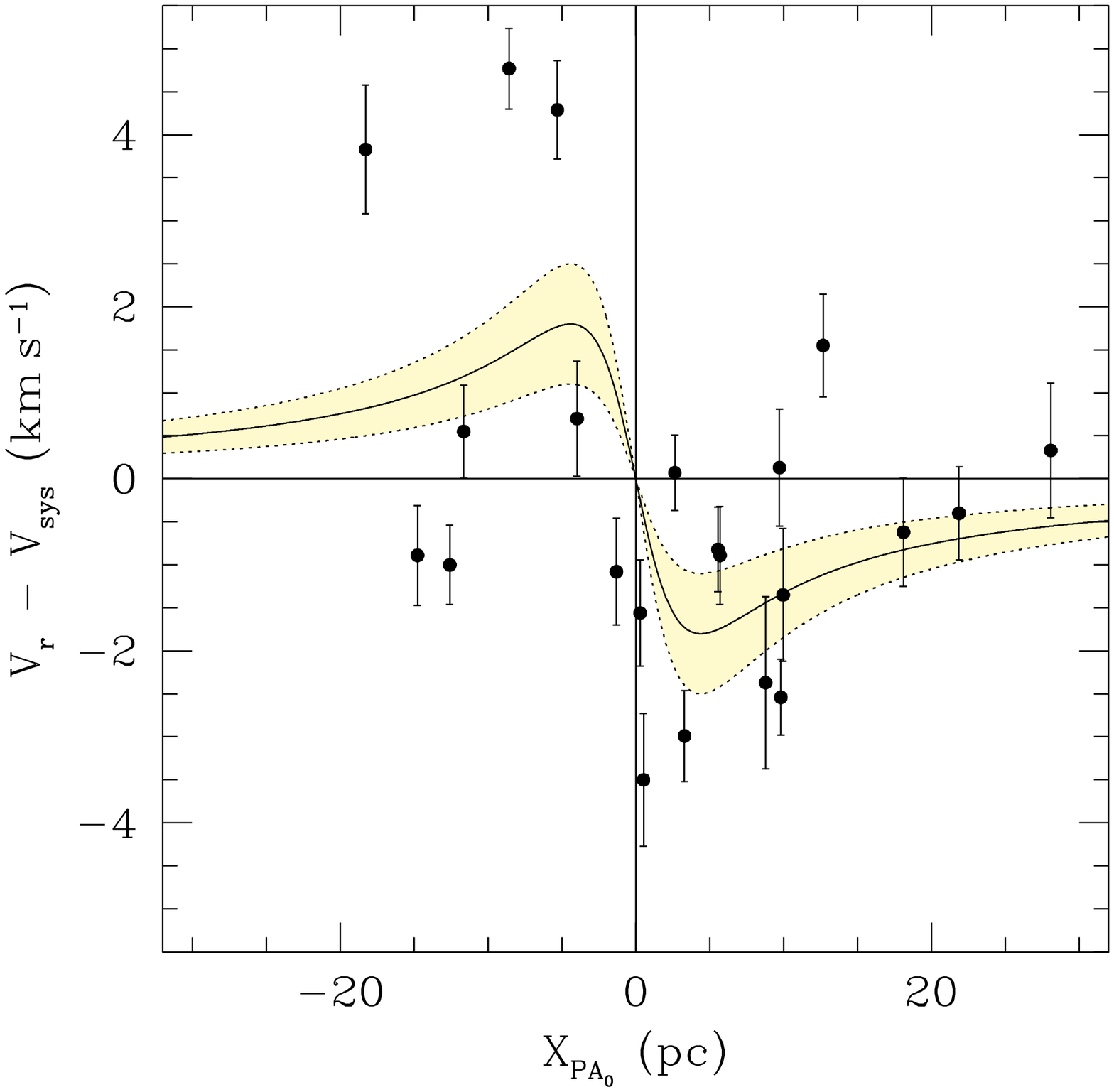}
\end{center}
\caption{Additional evidence for systemic rotation in NGC 1846. The upper panel shows a tangent plane
projection of the positions and velocities of the measured stars within the cluster. Red crossed points
(alternatively blue dotted points) indicate stars with velocities greater (less) than the global mean velocity.
The point sizes are proportional to the deviation of the velocity from the mean. The preferred axis of
rotation, oriented at a position angle $60 \pm 20\degr$ east of north, is marked. In the lower panel we
plot a rotation curve for NGC 1846, where $X_{\rm PA_0}$ is the perpendicular distance of from the rotation
axis and we have assumed an LMC distance modulus $\mu = 18.5$. The best fitting model, of
the form of Equation \ref{e:rotcurve} is marked, along with the curves represented by $\pm 1\sigma$ on
the maximum amplitude (see text).\label{f:rotcurve}}
\end{figure}
 
As a qualitative check on our detection of rotation in NGC 1846 we used the best-fit solution derived from
the full sample to construct a rotation curve. The top panel of Figure \ref{f:rotcurve} shows a tangent plane
projection of the positions and velocities of our measured members within the cluster. The preferred axis of
rotation, lying along the line PA$\;=60\degr-240\degr \pm 20\degr$, is indicated. In the lower panel we plot,
for each star, the offset from the cluster systemic velocity (derived in Section \ref{ss:meandisp} below) as a
function of the perpendicular distance from the rotation axis ($X_{\rm PA_0}$). Bearing in mind that almost all
of the point-to-point scatter can be accounted for by the cluster velocity dispersion (which is of order 
$2$\ km$\,$s$^{-1}$ -- again, see Section \ref{ss:meandisp} below), these
plots clearly support the detection of systemic rotation in NGC 1846. As an indicative measure, we fit a
rotation curve of the form \citep[as in e.g.,][]{lyndenbell:67,gott:73,henault:12a}:
\begin{equation} 
V_{\rm rot} = \frac{2 A_{\rm rot}}{r_{\rm peak}} \times \frac{X_{\rm PA_0}}{1 + (X_{\rm PA_0} / r_{\rm peak})^2}\,\,,
\label{e:rotcurve}
\end{equation}
where $r_{\rm peak}$ is the projected radius at which the peak amplitude of rotation occurs. The best fitting
model has $A_{\rm rot} = 1.8 \pm 0.7$\ km$\,$s$^{-1}$ and $r_{\rm peak} = 4.4 \pm 3.1$\ pc. Although not
particularly strongly constrained, these results are entirely consistent with those from our initial analysis above.

\subsection{Mean velocity and velocity dispersion}
\label{ss:meandisp}
Next, we corrected the velocity of each of our tracers for the systemic rotation, and calculated the global mean 
velocity and velocity dispersion, along with the velocity dispersion as a function of distance from the cluster center. 
To make the rotation 
corrections, we used the best fit to the full-sample rotation curve (i.e., the top panel of Figure \ref{f:rotate}). 
Although, as discussed, we suspect that the degree of rotation does vary as a function of radius within the 
cluster, our sample of kinematic probes is not sufficiently large to accurately measure this variation; 
the use of the mean rotation curve represents an adequate compromise. In any case, we note that even the extreme
case of applying no correction for rotation changes the global mean velocity by less than the 
uncertainty on the measurement, and the velocity dispersion results by less than $\sim 15\%$.

\begin{figure}
\includegraphics[width=86mm]{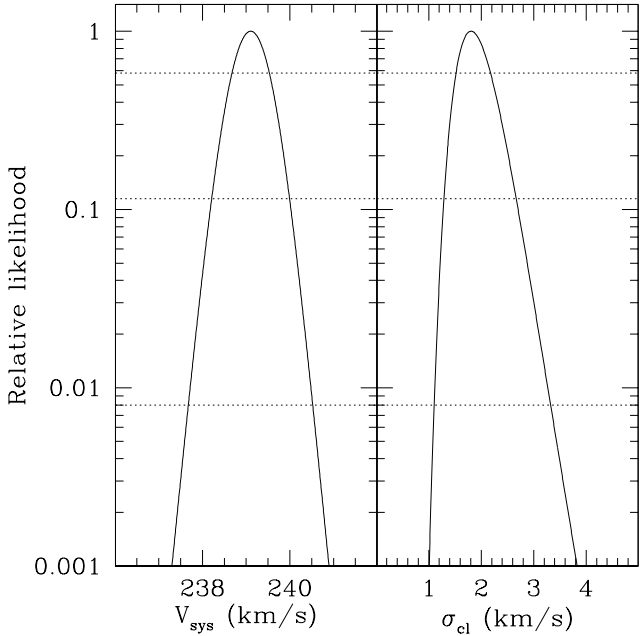}
\caption{Relative likelihood distributions for the systemic velocity $V_{\rm sys}$ and the mean velocity dispersion $\sigma_{cl}$, where in each case we have marginalized with respect to the other variable. The dashed horizontal lines indicate the levels corresponding to $1\sigma$, $2\sigma$, and $3\sigma$ uncertainties.\label{f:maxlike}}
\end{figure}

To calculate the mean radial velocity and velocity dispersion for NGC 1846 we used a maximum likelihood
technique following that defined by \citet{walker:06}. We assume that the measured velocities for
our stars, denoted $\{v_1 , \ldots , v_N\}$, are normally distributed about the systemic velocity $V_{\rm sys}$ according
to the associated measurement uncertainties $\sigma_i$ and the intrinsic cluster velocity dispersion $\sigma_{cl}$.
We can obtain numerical estimates for the quantities $V_{\rm sys}$ and $\sigma_{cl}$ by maximising the logarithm of
the joint probability function for $\{v_1 , \ldots , v_N\}$ -- i.e.,
\begin{equation}
\ln(p) = -\frac{1}{2} \sum_{i=1}^{N} \ln(\sigma_i^2 + \sigma_{cl}^2) -\frac{1}{2} \sum_{i=1}^{N} \frac{(v_i - V_{\rm sys})^2}{(\sigma_i^2 + \sigma_{cl}^2)} -\frac{1}{2} N \ln(2\pi)  .
\label{e:maxlike}
\end{equation}
Figure \ref{f:maxlike} shows the relative likelihood of both $V_{\rm sys}$ and $\sigma_{cl}$, where in each case
we have marginalized with respect to the other variable. To determine $1\sigma$ uncertainties we calculated the 
parameter limits for the region containing the central $68.3\%$ of the distribution function. 
%This procedure allowed us to account naturally for our sample size, as well as the different individual 
%uncertainties on our measured radial velocities.

We find a systemic velocity for NGC 1846 of $V_{\rm sys} = 239.1\pm 0.4$\ km$\,$s$^{-1}$. 
This is consistent with the value of $240\pm 10$\ km$\,$s$^{-1}$ determined by \citet{olszewski:91},
but slightly larger than that of $235.2\pm 0.9$\ km$\,$s$^{-1}$ from \citet{grocholski:06}.
For the mean velocity dispersion we find $\sigma_{cl} = 1.81^{+0.37}_{-0.29}$\ km$\,$s$^{-1}$. A value of zero
is excluded at $\gg 5\sigma$, indicating that we have resolved the intrinsic cluster dispersion.

The kinematic properties of globular clusters are commonly framed in terms of the central velocity dispersion,
$\sigma_0$. Given that we have members extending to $\approx 4 r_h$ it is likely that our measurement of
$\sigma_{cl}$ is an under-estimate of this quantity. To assess this we recalculated the velocity dispersion
as a function of cluster radius. Because of our relatively small sample size we are restricted to just three 
radial bins, corresponding roughly to stars inside $r_h$, stars between $r_h$ and twice $r_h$, and stars 
outside $2r_h$. For each bin we utilized Eq. \ref{e:maxlike} as before, but with fixed $V_{\rm sys} = 239.1$\ km$\,$s$^{-1}$.

\begin{figure}
\begin{center}
\includegraphics[width=86mm]{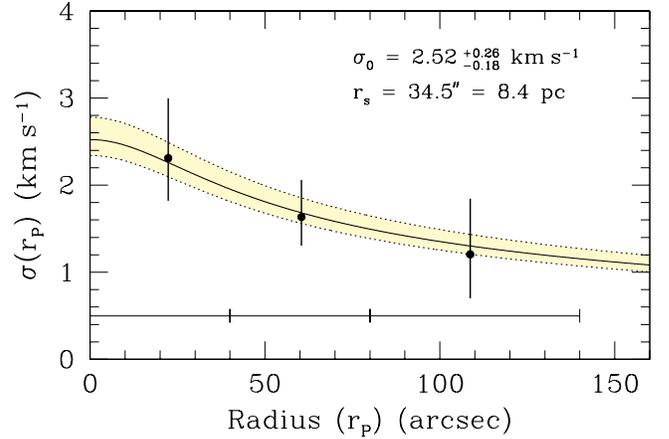}
\end{center}
\caption{Our indicative three-point velocity dispersion profile for NGC 1846, together with the best-fitting Plummer model described in the text. The shaded region corresponds to the $1\sigma$ uncertainties on the central dispersion $\sigma_0$.\label{f:dispersion}}
\end{figure}

Our results are shown in Figure \ref{f:dispersion}. A marginally significant decrease in the velocity dispersion 
with radius is evident. Even in the outermost bin the dispersion is resolved (i.e., a value of zero 
is excluded at a $\approx 3\sigma$ level). As an indicative measure we assume NGC 1846 is 
isotropic and fit a projected \citet{plummer:11} model to our three-point dispersion profile:
\begin{equation}
\sigma^2(r_p) = \frac{\sigma_0^2}{\sqrt{1+r_p^2/a^2}} . 
\label{e:plummer}
\end{equation}
Here $a$ is a scale radius which, for the family of projected \citet{plummer:11} models, is equal to the
half-light (half-mass) radius $r_h$ if the mass-to-light ratio $M/L$ is constant within the cluster. Note that we 
are not asserting here that a Plummer model is the most appropriate model to describe the internal kinematics of 
NGC 1846 -- with our poor radial resolution we are not in a position to undertake such an analysis. Rather, we 
use the Plummer model as a convenient parametrization to investigate how $\sigma_0$ relates to both our 
measurement of $\sigma_{cl}$ and our measure of $\sigma$ within $\approx r_h$, in a system with a 
constant-density core\footnote{Most
globular clusters are found to conform closely to models in which the density of stars within 
the cluster core is approximately constant. The family of \citet{king:62,king:66} models are the most prominent 
examples, but others such as those of \citet{elson:87} or \citet{wilson:75} are seen to provide superior 
fits in some cases \citep[e.g.,][]{mclaughlin:05}. The radial profile of \citet{goudfrooij:09} for NGC 1846 clearly
shows this cluster to be well described by models possessing constant-density cores.}.

In principle both $\sigma_0$ and $a$  are free parameters in Eq. \ref{e:plummer}. The degree by which the
scale radius $a$ differs from $r_h$ as determined from a surface-density or surface-brightness profile indicates 
how $M/L$ 
may vary with radius in the cluster \citep[see e.g.,][]{lane:10b}. Here, however, we again do not have a sufficiently
large sample of kinematic tracers, especially at inner radii, to put adequate constraints on $a$. We therefore
make the assumption that mass follows light (i.e., $a = r_h = 34.5\arcsec$) and fit for $\sigma_0$. The results of 
this process are seen in Figure \ref{f:dispersion}. We find $\sigma_0 = 2.52^{+0.26}_{-0.18}$\ km$\,$s$^{-1}$, only slightly 
larger than the value measured for our centralmost bin, but considerably in excess of our global dispersion. 

\subsection{Cluster mass \& luminosity}
\label{ss:masslum}
Having estimated the central velocity dispersion of NGC 1846 we can in turn derive an estimate of its
mass, since, for an isotropic Plummer model \citep[see e.g.,][]{dejonghe:87}:
\begin{equation}
M = \frac{64 a \sigma_0^2}{3 \pi G}\,\,.\vspace{1.5mm}
\label{e:plummer2}
\end{equation}
Substituting in our assumed value for $a = 8.4$\ pc and our estimate for $\sigma_0$,
we obtain $M = (8.4^{+1.7}_{-1.2})\times10^4\;{\rm M_\odot}$. 
%Note that \citet{goudfrooij:09} do not quote uncertainties on 
%the structural parameters they derive, so in calculating the uncertainty on $M$ we adopted a measurement error of 
%$\pm 10\%$ (which is conservative for LMC clusters observed with HST -- see e.g., \citet{mackey:03}).

We would also like to obtain an estimate of the global mass-to-light ratio for NGC 1846. To do this we first 
need an expression for the cluster luminosity, which we derive using Eq. \ref{e:king} integrated
with respect to $2\pi r_p dr_p$:
\begin{equation}
L(r_p) = \pi r_c^2 \Sigma_{V,0} \left[\ln\left(\alpha \right) - 4 \frac{\sqrt{\alpha} - 1}{\sqrt{\beta}} + \frac{\alpha - 1}{\beta} \right] \vspace{1mm}
\label{e:king2}
\end{equation}
where
\[
\alpha = 1 + (r_p / r_c)^2 \hspace{3mm} {\rm and} \hspace{3mm} \beta = 1 + (r_t / r_c)^2 \,\,.
\]
Here, $\Sigma_{V,0}$ is the $V$-band surface brightness corresponding to $n_0$ in Eq. \ref{e:king}. Note that 
in directly scaling the number-density profile of \citet{goudfrooij:09} to a surface-brightness profile we assume
that NGC 1846 is not strongly mass segregated. That it is a very diffuse cluster for its age
\citep[see e.g.,][]{mackey:08b,keller:11} is consistent with this assertion. Note that we also made this 
assumption previously when setting $a=r_h$ in Eq. \ref{e:plummer} in order to estimate $\sigma_0$.

To obtain a measurement of the total cluster $V$-band luminosity $L_V$ we set $r_p = r_t$ in Eq. \ref{e:king2} 
and estimated the central surface brightness $\Sigma_V(0)$ directly from our master ACS/WFC F555W reference 
image. To mitigate the effects of any mis-centering, and random fluctuations due to the brightest stars, we 
measured the flux within a relatively large aperture of radius $r_p=10\arcsec$. This is acceptable because the 
core radius of NGC 1846 is even larger still. We obtained $\Sigma_V(0) = 19.6\pm0.1$\ mag$\;$arcsec$^{-2}$, 
which, assuming the LMC distance modulus is $18.5$ and the foreground reddening $E(B-V) = 0.08$, corresponds 
to $\Sigma_V(0) = 670 \pm 62\;{\rm L_\odot}\;$pc$^{-2}$. However, we must also take into account that 
NGC 1846 is set against a moderately dense LMC field -- we measure the local surface brightness to be 
$\Sigma_{V,bkg}=22.8$\ mag$\;$arcsec$^{-2} = 35\;{\rm L_\odot}\;$pc$^{-2}$ by using regions on the ACS/WFC 
image beyond $r_t$. Thus, for the cluster only, $\Sigma_V(0) = 635 \pm 62\;{\rm L_\odot}\;$pc$^{-2}$ and,
substituting into Eq. \ref{e:king}, we find $\Sigma_{V,0} = 1.41 \Sigma_V(0) = 897 \pm 87\;{\rm L_\odot}\;$pc$^{-2}$.
Then, from Eq. \ref{e:king2}, $L_V = (1.44 \pm 0.14)\times10^5\;{\rm L_\odot}$.  

This leads to a global mass-to-light ratio $M / L_V = 0.59^{+0.13}_{-0.10}$ for NGC 1846. To place this value in a 
useful context, we compare it to the predictions of two well-known population synthesis codes -- the {\sc P\'{e}gase} 
(v2.0) models of \citet{fioc:97}, and the {\sc Galaxev} models of \citet{bruzual:03} -- for a cluster of the age 
and metal abundance of NGC 1846. For both sets of models we assume $Z=0.008$ and a single instantaneous 
burst of star formation. We further adopt the default initial stellar mass ranges --
$0.1-120\;{\rm M_\odot}$ for {\sc P\'{e}gase} and $0.1-100\;{\rm M_\odot}$ for {\sc Galaxev} -- and the
closest available initial mass function (IMF) to the \citet{kroupa:01} ``universal'' IMF. For the {\sc P\'{e}gase}
models this corresponds to the IMF of \citet{kroupa:93} while for the {\sc Galaxev} models it is the IMF of
\citet{chabrier:03}. The {\sc P\'{e}gase} models allow us to make the assumption that all white dwarfs are
retained in the cluster, while all neutron stars and black holes are expelled upon formation. For the
{\sc Galaxev} models we necessarily assume all stellar remnants remain in the cluster; the
predicted mass-to-light ratios reported below would drop by a few hundredths if the contributions from
neutron stars and black holes were excluded.

The results are as follows. The {\sc P\'{e}gase} models predict $M / L_V = 0.69$ at an age of $1.6$ Gyr and 
$M / L_V = 0.88$ at $2.0$ Gyr, while the {\sc Galaxev} models predict $M / L_V = 0.53$ and $0.70$
at these two ages. The differences between the two sets of results seem to be mainly due to the assumed 
IMFs and initial stellar mass ranges. Taken at face value and within the precision of our present measurements, 
the global mass-to-light ratio we have observed for NGC 1846 agrees acceptably with expectations derived purely 
from analysis of its constituent stellar populations. Note, however, that the systemic rotation 
we have detected could well play an important enough role in the internal kinematics of the cluster that the use 
of Eq. \ref{e:plummer2} when estimating $M / L_V$ may not be appropriate 
(see Sections \ref{ss:robustness} and \ref{s:discuss}).

\subsection{Relaxation time}
\label{ss:relax}
Finally, we take the parameters measured above and use them to derive the central and half-mass relaxation times
in NGC 1846. Following \citet{djorgovski:93}, the central relaxation time is given by:
\begin{equation}
t_{rc} = 8.338\times 10^6 \,{\rm yr}\; \times \frac{\rho_0^{1/2} r_c^3}{\bar{m} \ln\Lambda} \,\,, \vspace{1.5mm}
\label{e:centralrelax}
\end{equation}
where $\rho_0$ is the central mass density in the cluster, $\bar{m}$ is the mean stellar mass, and the
units of mass and distance are solar masses and parsecs, respectively. The quantity $\ln\Lambda$ is
the Coulomb logarithm where $\Lambda \approx 0.4N$ if $N$ is the total number of stars in the system.

Simple population synthesis models \citep[such as {\sc simclust},][]{deveikis:08} show that in a cluster
of age $\approx 1.75$ Gyr with a \citet{kroupa:01} mass function, the mean stellar mass 
$\bar{m} \approx 0.5\;{\rm M_\odot}$. Here we again assume that NGC 1846 is not strongly mass segregated
so that $\bar{m}$ does not vary significantly throughout the system. The number of stars in NGC 1846
is $N = M / \bar{m} \approx 2M$. To derive $\rho_0$ we used the two procedures outlined by \citet{djorgovski:93} 
to convert our observed, extinction-corrected central surface-brightness $\Sigma_V(0)$ to a central luminosity 
density $j_0 = 71 \pm 5\;{\rm L_\odot}\;$pc$^{-3}$, and then multiplied by our derived mass-to-light ratio so that 
$\rho_0 = 42^{+10}_{-8}\;{\rm M_\odot}\;$pc$^{-3}$. Substituting all these values into Eq. \ref{e:centralrelax} we
obtain a central relaxation time $t_{rc} = 2.4^{+0.3}_{-0.2}$\ Gyr. 

The half-mass, or median, relaxation time is given by \citep[e.g.,][]{binney:87}:
\begin{equation}
t_{rh} = 2.055\times 10^6 \,{\rm yr}\; \times \frac{M^{1/2} r_h^{3/2}}{\bar{m} \ln\Lambda} \,\,.  \vspace{1.5mm}
\label{e:medianrelax}
\end{equation}
As before, the units of mass and distance in this equation are solar masses and parsecs, respectively.
Application of the appropriate values for NGC 1846 yields $t_{rh} = 2.6^{+0.3}_{-0.2}$\ Gyr. This is only 
marginally longer than the central relaxation time, consistent with the observation that NGC 1846 is a diffuse cluster
(i.e., the core radius $r_c$ is comparable to $r_h$). 

\subsection{Robustness of the kinematic analysis}
\label{ss:robustness}
The magnitude of the rotation we infer in NGC 1846 is roughly comparable to the mean velocity 
dispersion in the system, which is quite unusual for a stellar cluster (see Section \ref{s:discuss}, below). 
Because of the potential significance of this observation, it is important to thoroughly assess the robustness 
of our kinematic analysis. There are two main reasons why it is necessary to 
undertake such a check. First, our sample of dynamical probes is comparatively small and so it is not 
inconceivable that, for example, a chance arrangement of velocity and position angle among a few stars 
could lead to a false rotation detection. Second, NGC 1846 has a non-neglibigle binary star fraction 
(note the strong binary star main sequences seen in the two CMDs in Figure \ref{f:cmds})
and so for a subset of stars in our ensemble the observed velocities likely possess extra components 
due to stellar companions rather than constituting a pure representation of the cluster kinematics.

To investigate these issues we developed $17$ sets of $10^5$ random realizations of our measured sample, 
and calculated for each set (i) how frequently we detected a rotation signal comparable to that observed for 
NGC 1846; and (ii) the mean ensemble velocity dispersion. We characterized a given set of models according 
to the assumed properties of its binary star population. This analysis is quite similar, at least in its
principle, to that of \citet{mcconnachie:10}.

Each realization consisted of $22$ targets at identical radii to our real sample, but at random position angles
and with random velocities generated according to a well-defined set of rules. It could be argued
that the radius assigned to each star should also be randomly selected, according to the density profile
of NGC 1846. However, our measured sample is not solely defined by this density profile -- rather it is the
profile convolved with some complex and essentially unknown selection function for input to FLAMES
(see Section \ref{ss:targets}). In the absence of this information, we felt the most sensible option was to 
maintain the radial distribution of the observed sample for each mock system.

For each target we first assigned a base velocity by selecting randomly from a Gaussian distribution of 
width specified by Equation \ref{e:plummer} with $a = r_h = 34.5\arcsec$ and $\sigma_0 = 2.5$\ km$\,$s$^{-1}$, 
evaluated at the appropriate radius. To this we added a random Gaussian deviate to represent the observational
uncertainty, exactly as in Section \ref{ss:rotation}. Next, we randomly assigned the star to be an
unresolved binary, or not, according to the assumed binary fraction for the overall set to which the
mock realization belonged. If a star was selected to be a binary we randomly generated a set of parameters 
to characterize the system, again according to the properties of the overall set, leading to an additional
component to add to the base velocity. If, however, a star was not selected to be a binary, 
no further modification of its base velocity was made. Note that the planetary nebula Mo-17 was never 
selected to be a binary (since our velocity for this object comes from the nebula itself rather than the central star), 
but was otherwise treated identically to the other $21$ stars in the sample.

Binary systems are characterized by the masses of the two components, $m_1$ and $m_2$ (where the
mass of the primary $m_1 \ge m_2$), the orbital period $P$, and the orbital eccentricity $e$. It is
convenient to define the mass ratio $q = m_2 / m_1$. The semi-major axis $a_2$ of the orbit of the
secondary about the barycenter may be calculated using Kepler's third law, and the semi-major axis
of the primary $a_1 = qa_2$. To place a given binary system into the observational plane requires
several additional parameters -- the inclination $i$ of the orbit to the line of sight, the argument
(or longitude) or periastron, $\omega$, and the orbital phase $\theta$ at which the observation was
made. The radial velocity of the primary then varies as:\vspace{1.5mm}
\begin{equation}
V_{b,1} = \frac{ 2\pi a_1 \sin i}{P \sqrt{1-e^2}} \left[ \cos(\theta + \omega) + e\cos(\omega) \right]\,\,. \vspace{2mm}
\label{e:binvel}
\end{equation}

We defined each set of random realizations according to the distributions from which the mass
ratio, period, and eccentricity of each binary were drawn. Mass ratios were selected from
either a uniform distribution or the normal distribution of \citet{duquennoy:91}, while periods 
were chosen according to either a uniform distribution in $\log P$ or the log-normal distribution
of \citet{duquennoy:91}. Orbital eccentricities were selected from one of four distributions:
circular orbits only ($e = 0$), a uniform distribution, the normal distribution of \citet{duquennoy:91},
or the thermal distribution of \citet{heggie:75}. The different combinations of these 
distributions defined $16$ sets of models. We also computed a seventeenth set with no binaries
for comparison purposes.

\begin{deluxetable*}{ccccccccccccccc}
\tabletypesize{\scriptsize}
\tablecaption{Results from the mock realizations of our measured ensemble, including the effects of unresolved binary stars.\label{t:mocks}}
\tablehead{
\colhead{Set} & \multicolumn{4}{c}{Assumed distributions} & \multicolumn{3}{c}{Peak$\,\ge 1.1\,$km$\,$s$^{-1}$ (\%)} & \multicolumn{3}{c}{Peak$\,\ge 0.7\,$km$\,$s$^{-1}$ (\%)} & \multicolumn{4}{c}{Mean kinematics (km$\,$s$^{-1}$)}\vspace{1mm}\\
\colhead{Number} & \colhead{$q$} & \colhead{$\log P$} & \colhead{$e$} & \colhead{$f_b$} & \colhead{$A_{\rm rot}$} & \colhead{$+\,$rms} & \colhead{$+R^2$} & \colhead{$A_{\rm rot}$} & \colhead{$+\,$rms} & \colhead{$+R^2$} & \colhead{$\bar{V}_{\rm sys}$} & \colhead{$\sigma_{V_{\rm sys}}$} & \colhead{$\bar{\sigma}_{cl}$} & \colhead{$\sigma_{\sigma_{cl}}$}
}
\startdata
01 & $...$ & $...$ & $...$ & $0.00$ & $2.01$ & $0.23$ & $0.39$ & $8.88$ & $1.31$ & $0.97$ & $239.10$ & $0.41$ & $1.79$ & $0.32$\vspace{1mm} \\
02 & Unif. & Unif. & Circ. & $0.33$ & $6.28$ & $0.33$ & $0.72$ & $18.15$ & $1.35$ & $1.30$ & $239.11$ & $0.51$ & $2.18$ & $0.42$ \\
03 & Unif. & Unif. & Unif. & $0.33$ & $5.98$ & $0.34$ & $0.74$ & $17.80$ & $1.34$ & $1.23$ & $239.09$ & $0.50$ & $2.18$ & $0.42$ \\
04 & Unif. & Unif. & Ther. & $0.33$ & $5.93$ & $0.32$ & $0.75$ & $17.66$ & $1.32$ & $1.26$ & $239.11$ & $0.50$ & $2.16$ & $0.43$ \\
05 & Unif. & Unif. & Norm. & $0.33$ & $6.20$ & $0.33$ & $0.73$ & $17.96$ & $1.26$ & $1.23$ & $239.09$ & $0.50$ & $2.18$ & $0.42$\vspace{1mm} \\
06 & Unif. & Norm. & Circ. & $0.33$ & $6.01$ & $0.33$ & $0.69$ & $18.10$ & $1.35$ & $1.26$ & $239.11$ & $0.50$ & $2.16$ & $0.42$ \\
07 & Unif. & Norm. & Unif. & $0.33$ & $5.98$ & $0.31$ & $0.67$ & $17.58$ & $1.38$ & $1.26$ & $239.10$ & $0.50$ & $2.17$ & $0.42$ \\
08 & Unif. & Norm. & Ther. & $0.33$ & $5.86$ & $0.38$ & $0.76$ & $17.44$ & $1.30$ & $1.25$ & $239.10$ & $0.50$ & $2.16$ & $0.41$ \\
09 & Unif. & Norm. & Norm. & $0.33$ & $5.76$ & $0.31$ & $0.75$ & $17.61$ & $1.29$ & $1.25$ & $239.10$ & $0.50$ & $2.17$ & $0.42$\vspace{1mm} \\
10 & Norm. & Unif. & Circ. & $0.55$ & $8.21$ & $0.32$ & $0.90$ & $21.73$ & $1.21$ & $1.40$ & $239.10$ & $0.53$ & $2.33$ & $0.43$ \\
11 & Norm. & Unif. & Unif. & $0.55$ & $8.24$ & $0.33$ & $0.86$ & $21.68$ & $1.25$ & $1.34$ & $239.09$ & $0.53$ & $2.33$ & $0.43$ \\
12 & Norm. & Unif. & Ther. & $0.55$ & $8.07$ & $0.33$ & $0.89$ & $21.58$ & $1.21$ & $1.38$ & $239.10$ & $0.54$ & $2.33$ & $0.43$ \\
13 & Norm. & Unif. & Norm. & $0.55$ & $8.23$ & $0.38$ & $0.88$ & $21.72$ & $1.25$ & $1.33$ & $239.10$ & $0.54$ & $2.32$ & $0.42$\vspace{1mm} \\
14 & Norm. & Norm. & Circ. & $0.55$ & $7.96$ & $0.34$ & $0.81$ & $21.14$ & $1.24$ & $1.30$ & $239.09$ & $0.54$ & $2.30$ & $0.42$ \\
15 & Norm. & Norm. & Unif. & $0.55$ & $7.78$ & $0.32$ & $0.89$ & $21.19$ & $1.26$ & $1.40$ & $239.10$ & $0.52$ & $2.29$ & $0.43$ \\
16 & Norm. & Norm. & Ther. & $0.55$ & $7.80$ & $0.36$ & $0.90$ & $20.96$ & $1.28$ & $1.39$ & $239.09$ & $0.53$ & $2.31$ & $0.42$ \\
17 & Norm. & Norm. & Norm. & $0.55$ & $7.79$ & $0.29$ & $0.80$ & $20.93$ & $1.22$ & $1.31$ & $239.10$ & $0.53$ & $2.29$ & $0.42$
\enddata
\end{deluxetable*}
 
Since all our observed targets are upper RGB stars (with a few possible AGB stars), for every mock binary 
system we set the primary mass $m_1 = 1.625 M_\odot$, based on the typical masses of upper RGB stars
in the best-fitting Dartmouth isochrones \citep{dotter:08} for NGC 1846 
\citep{goudfrooij:09}. The binary fraction for a given set of models was defined
using the observations of \citet{milone:09} extrapolated according to the assumed distribution
of mass ratios. Milone et al.\ estimated the binary fractions for $12$ intermediate-age
LMC clusters from HST/ACS imaging by counting stars above the main sequence. Due to
blends and photometric uncertainties they were only able to clearly identify binaries with
$q \ge 0.6$ or $0.7$, depending on the quality of the imaging. They found typical binary 
fractions $f_b \sim 0.14$ for $q \ge 0.6$, or $f_b \sim 0.09$ for $q \ge 0.7$. For a
a uniform distribution in $q$ these correspond to a total binary fraction $f_b \approx 0.33$,
while for the \citet{duquennoy:91} normal distribution they correspond to $f_b \approx 0.55$.
We adopted these two values for our various sets of models, depending on the assumed
distribution in $q$.

Having selected a given star to be a binary, our procedure for generating a radial velocity was
as follows. We first randomly selected a mass ratio, period, and eccentricity from the relevant
distributions. We defined a lower bound to allowed values of $q$ by noting that the mass limit 
for hydrogen-burning, $m_2 \approx 0.08 M_\odot$, corresponds to $q \approx 0.05$ in our case.
We further defined limits to the allowed values of $P$ following the procedure outlined by 
\citet{mcconnachie:10}. Briefly, the lower limit is set by the minimum orbital separation before 
the onset of mass-transfer, while the upper limit is set by the orbital separation corresponding 
to the boundary between ``hard'' and ``soft'' binaries in NGC 1846 -- the latter of which are
efficiently destroyed by three- or four-body interactions within the cluster.
Next, we determined the orientation of the binary by selecting the angles
$i$ and $\omega$ from uniform distributions. Finally, we selected a number in the
range $0-1$, also according to a uniform distribution, to represent the elapsed fraction of 
the binary's orbital period since periastron, and converted this into the phase $\theta$ by numerically 
solving Kepler's equation. With all the necessary parameters in hand, we determined the radial
velocity according to Eq. \ref{e:binvel}. To mimic our original identification of cluster members 
(Section \ref{ss:memstar}), if the resulting velocity of the star was separated from the systemic 
cluster velocity by more than five times the velocity dispersion at that radius, we returned and 
regenerated all its binary parameters.
 
The results for all $17$ sets of realizations are presented in Table \ref{t:mocks}. To illustrate
our analysis procedure we consider set 1 -- the control set with zero binary fraction. We quantified 
whether the inferred rotation of a mock system matched that observed for the real cluster,
by using the amplitude of rotation and two quality of fit parameters. As described in Section
\ref{ss:rotation}, the rotation velocity implied by the NGC 1846 curve shown in the top panel of Figure 
\ref{f:rotate} is $1.1 \pm 0.4$\ km$\,$s$^{-1}$. The curve is, further, a high quality sinusoid, 
suggesting that the inferred rotation is not simply due to stochastic deviations from zero at a few 
position angles. The rms residual for the best-fitting model is $0.27$, and the coefficient of 
determination $R^2 = 0.97$. 

Just $2.01\%$ of our mock systems in set 1 had inferred rotation of matching or greater amplitude
than that measured for NGC 1846; however only about a tenth of these ($0.23\%$ of all systems) also 
had an rms residual value equal to or smaller than the best-fit model for NGC 1846. A slightly larger
fraction ($0.39\%$ of all systems) had matching or greater rotation amplitude together with an $R^2$ 
value equal to or greater than our best-fit model.
Considering the lower bound on our NGC 1846 rotation amplitude of $\sim 0.7$\ km$\,$s$^{-1}$, 
defined by the uncertaintly on the measurement, the fraction of mock systems with matching or
greater amplitude rose to $8.88\%$. Again, however, only a small subset of these also had an
equivalently high-quality rms value ($1.31\%$ of all systems) or $R^2$ value ($0.97\%$ of all systems).

To give a visual indication of the goodness-of-fit criteria, we show in Figure \ref{f:mocks} several
examples (taken from set 17, see below) where a rotation amplitude greater than that measured for 
NGC 1846 is inferred, but where the rms and/or the $R^2$ values indicate a poor quality sinusoid. 
It is quite clear that these systems would not be mistaken for one in which a strong rotation signal 
had been reliably detected. We also show one example where stochastic fluctuations have resulted in a 
rotation curve that passess all our tests, and would have led to a false positive detection of
rotation in the cluster. Fortunately, as outlined in Table \ref{t:mocks}, this is a rare occurrence.

For each random realization of our sample we also calculated the mean velocity dispersion
using the maximum likelihood technique described previously. We then found the average dispersion
across all random realizations in a given set, along with the standard deviation in this value.
The mean systemic velocity for each random realization is a natural by-product of the maximum
likelihood calculation, and for completeness we determined the average value for this quantity across 
each set as well. The results of this process are visible in Table \ref{t:mocks}. For the control set 1,
we recover an average systemic velocity and velocity dispersion precisely matching those values 
derived for NGC 1846, including the uncertainties. This is to be expected, since the control run 
is the trivial case of re-measuring systems generated according to the velocity dispersion profile
seen in Figure \ref{f:dispersion}, with no contribution from binary stars.

We now consider the results for sets $2-17$, which possess binary population
characteristics as outlined in Table \ref{t:mocks}. As expected, the additional velocity components
due to the orbital motions of binary pairs do have some influence on our kinematic measurements.
However, the effects do not appear to be strongly dependent on the adopted distributions for
the orbital period or eccentricity. Changing the distribution for the mass ratio, $q$, has a more
significant effect; however this is most likely due to the resulting change in the inferred binary
fraction in the cluster -- more binaries imply a stronger influence on the kinematic measurements.

In terms of the inferred rotation, an increased percentage of the mock systems have amplitudes of matching
or greater amplitude than our NGC 1846 measurement. For $f_b = 0.33$ these values lie in the 
range $\sim 5.7-6.3\%$, while for $f_b = 0.55$ they are higher, $\sim 7.7-8.3\%$. As before, however,
only a very small subset of these systems possess high quality sinusoidal rotation curves -- the overall
fractions are less than $1\%$ for both goodness-of-fit criteria in all cases. Once again considering
the conservative case defined by $A_{\rm rot} \sim 0.7$\ km$\,$s$^{-1}$, the percentages all rise; however 
the overall fraction of systems in each set with a high amplitude and 
high-quality rotation curve is again always close to or below $1.5\%$.

The mean kinematics of each set of models are interesting. Binaries have no effect on the average systemic velocity,
which always comes out at the measured NGC 1846 value. This is because the extra binary component of 
an individual radial velocity is equally likely to be pointing towards, or away from, the observer. However 
the binary stars do appreciably inflate the mean cluster velocity dispersion. In the case where $f_b = 0.33$
the mean dispersion goes from $1.8$\ km$\,$s$^{-1}$ to nearly $2.2$\ km$\,$s$^{-1}$, while for $f_b = 0.55$
the dispersion is typically just above $2.3$\ km$\,$s$^{-1}$. As with the rotation measures, these values are
not strongly dependent on the adopted distributions for the orbital period or eccentricity.

\begin{figure}
\begin{center}
\includegraphics[width=86mm]{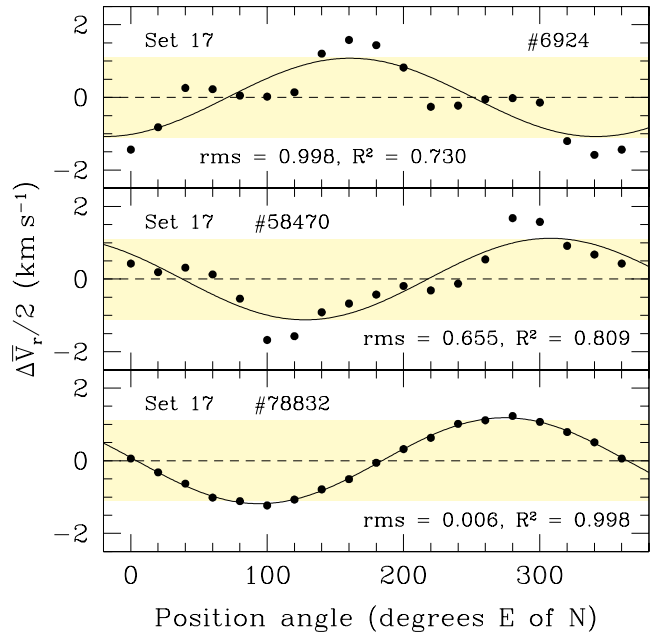}
\end{center}
\caption{Example rotation curves for three mock systems in set 17. We selected these to have comparable amplitudes to that measured for NGC 1846 (indicated, as previously, by the shaded region). The upper two panels show systems for which the best-fit models are of lower quality than than for NGC 1846, as defined by the rms of the fit and the coefficient of determination $R^2$ (see text). In the lower panel we show one rare example of a high quality curve arising purely by chance. This particular realization had the best quality fit of all $10^5$ mock systems in set 17.\label{f:mocks}}
\end{figure}
 
In summary, we draw the following conclusions from our Monte Carlo simulations. First, while
it is not impossible for rotation signals comparable to that which we measured for NGC 1846 to arise
stochastically, it is a very unusual occurrence with $< 1\%$ of systems in any given 
set exhibiting such characteristics. The presence of unresolved binary stars increases the number of 
systems for which a high amplitude of rotation could be inferred; however the rotation curves for such systems 
would invariably not be mistaken for the expected sinusoidal variation of velocity with position angle. 
Thus, while we cannot rule out that the rotation we have observed in NGC 1846 is simply down to
statistical fluke, we are better than $99\%$ confident that it is a genuine signal.

Second, it is likely that our measured velocity dispersion for NGC 1846, $\sigma_{cl}$, along with our 
inferred central velocity dispersion $\sigma_0$, are both inflated above their true values due to the
presence of unresolved binary stars in our sample. In our simulations, an input value of 
$\sigma_{cl} = 1.81$\ km$\,$s$^{-1}$ is increased to at most $\approx 2.33$\ km$\,$s$^{-1}$.
Since the various sources of dispersion add in quadrature, this implies the true $\sigma_{cl}$
could be as low as $\approx 1.10$\ km$\,$s$^{-1}$, and the true $\sigma_0$ as low as 
$\approx 2.05$\ km$\,$s$^{-1}$. By comparison, our typical measurement uncertainty of
$\approx 0.6$\ km$\,$s$^{-1}$ on each individual velocity inflates the true dispersions by
less than $0.1$\ km$\,$s$^{-1}$ (note that this contribution is accounted for by the
maximum likelihood technique we employed to measure the cluster dispersion).

\section{Discussion}
\label{s:discuss}
The key point of interest arising from our analysis of the internal kinematics of NGC 1846 concerns the
degree of systemic rotation, which appears high compared with, for example, many Milky Way globular 
clusters, or indeed the few Magellanic Cloud clusters for which such measurements exist. We find a 
ratio $A_{\rm rot} / \sigma_0 = 0.44 \pm 0.16$ if we adopt the mean amplitude of rotation
$A_{\rm rot} = 1.1 \pm 0.4$\ km$\,$s$^{-1}$. However, we also demonstrated that the maximum amplitude
of rotation in NGC 1846 might well be as high as $\approx 2$\ km$\,$s$^{-1}$ (especially since there
is the additional unknown factor $\sin i$ required to deproject our rotation measurement -- note that the
mean value of $\sin i$ assuming a uniform distribution of inclination angles is $2/\pi$). In this case,
$A_{\rm rot} / \sigma_0 \sim 0.8$. We also showed that unresolved binary stars in our sample probably
inflate the measured velocity dispersion by up to $\approx 0.5$\ km$\,$s$^{-1}$, implying that the
fraction of ordered motion with respect to pressure support is likely to be even larger still, 
$A_{\rm rot} / \sigma_0 \ga 1.0$. 

These latter two estimates are commensurate with the highest ratios 
measured in the sample of $24$ Galactic globular clusters compiled by \citet{bellazzini:12}; indeed 
even our lower limit of $A_{\rm rot} / \sigma_0 = 0.44$ would place NGC 1846 in the top $25\%$ of this 
sample. They are also somewhat larger than the values measured for young Magellanic Cloud clusters.
\citet{fischer:92a} found $A_{\rm rot} / \sigma_0 = 0.45 \pm 0.20$ in the $\approx 100$ Myr old
cluster NGC 1866, while \citet{fischer:93} measured a comparable degree of rotation in the 
$\approx 50$ Myr old cluster NGC 1850. More recently, \citet{henault:12a,henault:12b} measured
$A_{\rm rot} / \sigma_0 = 0.60 \pm 0.30$ for the very young ($\approx 3$ Myr) massive 
cluster R136 at the centre of 30 Doradus. Notably, however, the only other intermediate-age
LMC cluster investigated on a star-by-star basis, NGC 1978, shows no significant evidence
for rotation \citep{fischer:92b}.

When comparing the degree of internal rotation in NGC 1846 with rotation seen in Galactic 
globular clusters, it is important to bear in mind that NGC 1846 is still dynamically quite young.
As demonstrated in Section \ref{ss:relax}, its 
central and median relaxation times are greater than the cluster age. It is possible, even likely, that these 
time-scales were much shorter early on in the cluster's evolution \cite[see e.g.,][]{mackey:07b,mackey:08b}; 
however it is clear that in the ancient Galactic globular clusters any ordered motions have had far more 
opportunity to become randomized, thus surpressing $A_{\rm rot} / \sigma_0$ relative to NGC 1846.

This is not the case for the young LMC clusters NGC 1850 and 1866, and R136, which, like
NGC 1846, should be dynamically unevolved \citep[e.g.,][]{fischer:92a,fischer:93,henault:12a}.
That the degree of rotation observed in NGC 1846 is, arguably, greater than in these young systems
is a striking result. If, as we suspect, $A_{\rm rot} / \sigma_0 \approx 0.8-1.0$ then ordered
rotation is comparable in importance to random motions in providing support against gravitational
collapse. Following the discussion in \citet{henault:12a}, for this range in $A_{\rm rot} / \sigma_0 $,
between $\approx 35-45\%$ of the total cluster kinetic energy would be in rotation.

NGC 1846 is only mildly elliptical -- \citet{goudfrooij:09} measure $\epsilon = 0.12 \pm 0.02$.
However, a cluster's ellipticity is not necessarily a good indication of its degree of internal rotation.
\citet{bellazzini:12} found no correlation between these two quantities in their sample of Galactic 
globular clusters. Furthermore, the lone globular cluster associated with the isolated Local group 
dwarf irregular galaxy WLM, shows no evidence for strong internal rotation despite displaying
a high degree of ellipticity \citep{stephens:06}. NGC 1978, for which \citet{fischer:92b} found no
evident signs of rotation, also has a very flattened shape with $\epsilon \approx 0.3$.

\citet{bellazzini:12} demonstrated that for Galactic globular clusters $A_{\rm rot} / \sigma_0$
correlates quite strongly with $[$Fe$/$H$]$, such that more metal-rich systems have stronger internal
rotation. With $[$Fe$/$H$] \approx -0.4$, NGC 1846 fits this correlation well, although we again
caution that we are comparing dynamically evolved systems with a dynamically young system.
The suggestion made by \citet{bellazzini:12} is that the observed correlation between $[$Fe$/$H$]$ 
and $A_{\rm rot} / \sigma_0$ may hint that dissipative gas dynamics plays a significant role in the 
process of cluster formation, due to the fact that a larger metal content in a gas should imply a higher 
efficiency in energy dissipation via atomic transitions.

This idea is particularly relevant to the present case, where we strongly suspect the presence of 
multiple stellar generations in NGC 1846. \citet{bekki:10,bekki:11} uses hydrodynamic simulations 
to investigate the formation of a second generation of stars at the centre of a globular cluster following 
the accretion of gas (including AGB ejecta) into its potential well \citep[see also][]{bekki:09}. He finds 
that if the first generation of stars possesses even a very small net angular momentum, gaseous 
dissipation during accretion onto the cluster center leads to a dynamically cold, rapidly rotating 
second generation with $A_{\rm rot} / \sigma_0 \ga 0.8$. The second generation is initially very
centrally concentrated but this nested structure, and the strong rotational signature, ought to be
smoothed out as a result of relaxation processes during the subsequent long term evolution of the cluster.
Since NGC 1846 is, dynamically speaking, still quite young, both the central concentration
of the younger generation \citep[as measured by][]{goudfrooij:09}, and the rotational kinematics (as seen
in the present work), are apparently still evident. Our possible detection of increasing rotation
towards the cluster center is also consistent with this picture.

A more sensitive test of the model described above would be
achieved if we could split our kinematic sample into earlier and later generations; however, to
presage the results of our chemical abundance analysis somewhat (see Paper II), there is no obvious
marker for achieving this -- and in any case the present ensemble is probably too small for 
useful subdivision. The most reliable method would be to target a large number of stars across the EMSTO
in order to directly correlate kinematics against position on the CMD and within the cluster. 
Such measurements, if at all possible, would require a considerable investment of telescope time but 
would provide critical information to test the different formation hypotheses.

Nonetheless, the fact that we have detected strong rotation in the first EMSTO cluster subjected to 
detailed dynamical examination is suggestive that this may be an important feature of these systems.
In future it will be critically important to enlarge the sample of intermediate-age Magellanic Cloud
clusters for which internal kinematics have been measured, including both systems with and without 
an EMSTO. Realistic $N$-body modelling of clusters such as those predicted by hydrodynamic
simulations, in which there is a centrally-concentrated, strongly-rotating second generation of stars
embedded in a more diffuse, pressure supported first generation, would also be extremely 
useful (although we recognize the current limitations placed on such models due to the 
maximum particle number of $\sim 1-2 \times 10^5$). An issue of particular interest is
how the spatial and kinematic distinctions between the different generations propagate through the
dynamical evolution of the system, especially where the cluster is strongly mass segregated at early times
such that violent relaxation due to rapid stellar mass-loss drives significant expansion, as
appears to be necessary for the intermediate-age EMSTO clusters \citep[e.g.,][]{keller:11}. 

\section{Summary \& Conclusions}
In this paper we have described a detailed set of VLT/FLAMES observations of red giant stars in the
peculiar intermediate-age LMC star cluster NGC 1846, along with the data reduction procedure
we employed to extract and process individual spectra. In total, we targeted $29$ stars within
the nominal boundary of NGC 1846, of which $21$ possess radial velocities indicating
their membership of the cluster at high confidence. In addition, we targeted the planetary nebula 
Mo-17, and the radial velocity of this object indicates that it too is a member of the cluster.

We have used our spectra to investigate the elemental abundance patterns present in NGC 1846,
including the possibility of star-to-star variations in light element abundances. These results
will be presented in a forthcoming work (Paper II). In the present paper we took the radial velocity 
measurements for our sample and used these to conduct a thorough analysis of the internal
kinematics of the cluster, with the following results:
\begin{itemize}[leftmargin=*]
\item{NGC 1846 exhibits a significant degree of systemic rotation. The mean amplitude
is $A_{\rm rot} = 1.1 \pm 0.4$\ km$\,$s$^{-1}$, with the rotation axis oriented at $60 \pm 20\degr$
east of north. There are indications that
the rotation signal may vary with position in the cluster such that the amplitude increases
towards the center and peaks somewhere within the half-light radius. The maximum amplitude 
may well be as high as $A_{\rm rot} \approx 2$\ km$\,$s$^{-1}$,
especially considering that there is also a correction factor $\sin i$ for the unknown inclination
of the rotation axis to the line of sight. An extensive suite of Monte Carlo models suggests that,
because of the relatively small size of our sample and the presence of a significant population of
unresolved binary stars in NGC 1846, stochastic fluctuations could reproduce the observed rotation 
curve; however this only occurs very rarely -- less than $\approx 0.3\%$ of the time if no binaries
are present, or less than $\approx 1\%$ of the time for a cluster binary fraction of up to $f_b = 0.55$.}
\item{We measure a mean velocity dispersion $\sigma_{cl} = 1.81^{+0.37}_{-0.29}$\ km$\,$s$^{-1}$.
Assuming a simple parametrization of the velocity dispersion fall-off with radius, the implied
central velocity dispersion in the cluster is $\sigma_0 = 2.52^{+0.26}_{-0.18}$\ km$\,$s$^{-1}$.
Our Monte Carlo modeling suggests that the presence of unresolved binary stars in our sample
could substantially inflate these quantities. If the binary fraction $f_b = 0.55$, the true values
could be as low as $\sigma_{cl} \approx 1.1$\ km$\,$s$^{-1}$ and $\sigma_0 \approx 2.0$\ km$\,$s$^{-1}$.}
\item{The ratio of ordered motion to pressure support is formally 
$A_{\rm rot} / \sigma_0 = 0.44 \pm 0.16$; however, accounting for the probable maximum amplitude
of rotation in the cluster, the inclination factor $\sin i$, and the contribution of binary stars to inflating 
the observed velocity dispersion, this quantity is likely to be as high as $0.8-1.0$. In this case, between
$\approx 35-45\%$ of the total cluster kinetic energy would be in rotation.}
\item{Under the assumption that mass follows light in the cluster, the mass of NGC 1846 is
$(8.4^{+1.7}_{-1.2}) \times 10^4\, {\rm M}_\odot$ and the implied mass-to-light ratio is $0.59^{+0.13}_{-0.10}$,
consistent with predictions made purely on consideration of its consitutent stellar populations.
Note, however, that these quantities are derived assuming a ``dispersion-only'' cluster
(Eq. \ref{e:plummer2}). If internal rotation provides an important contribution against gravitational
collapse, as seems probable, more sophisticated modelling will be required to obtain reliable estimates
of mass and $M / L_V$.}
\item{The median relaxation time for NGC 1846 is $t_{rh} = 2.6^{+0.3}_{-0.2}$ Gyr, indicating that
the cluster is dynamically youthful. Hence any kinematic signatures encoded during its formation
ought to remain present.}
\end{itemize}

The observation that substantial rotation is present in NGC 1846, at a magnitude comparable
to that of the velocity dispersion, is consistent with the predictions of simulations 
modeling the formation of multiple generations in globular clusters \citep[see e.g.,][]{bekki:10,bekki:11}. 
It would be of significant interest to improve our knowledge of the internal kinematics of 
this cluster by extending the present work to a much larger sample, ideally one in which
the multiple generations could be easily identified. Similarly, by extending
our analysis to additional intermediate-age Magellanic Cloud clusters, both with and without
the extended main-sequence turn-off morphology, we could hope to learn whether strong
internal rotation is a key signature of the formation of clusters with multiple constituent
stellar populations.

\acknowledgements
We would like to thank the anonymous referee for their assistance in improving this paper, along with Mike Irwin 
and Vanessa Hill for helpful discussions at the commencement of the project. ADM is grateful for financial 
support from the Australian Research Council in the form 
of an Australian Research Fellowship (Discovery Projects grant DP1093431). GDaC is grateful for 
research support from the Australian Research Council through Discovery Projects grant DP120101237.
AMNF and ADM acknowledge support by a Marie Curie Excellence Grant from the European 
Commission under contract MCEXT-CT-2005-025869 during the early stages of this project. 

This work is partially based on observations made with the NASA/ESA Hubble Space Telescope, 
obtained from the Data Archive at the Space Telescope Science Institute, which is operated by 
the Association of Universities for Research in Astronomy, Inc., under NASA contract NAS 5-26555. 
These observations are associated with programs \#9891 and \#10595. 

This publication makes use of data products from the Two Micron All Sky Survey (2MASS), which 
is a joint project of the University of Massachusetts and the Infrared Processing and Analysis 
Center/California Institute of Technology, funded by the National Aeronautics and Space 
Administration and the National Science Foundation. \vspace{3mm}
%VALD?

{\it Facilities:} \facility{VLT:Kueyen (FLAMES/GIRAFFE)}.

\end{document}